\documentclass[%
superscriptaddress,
twocolumn,
 amsmath,amssymb,
 aps,
 superscriptaddress,
 pra,
]{revtex4-2}

\usepackage[T1]{fontenc}
\usepackage[utf8]{inputenc}
\usepackage{graphicx}%
\usepackage{dcolumn}%
\usepackage{bm}%
\usepackage{times}
\usepackage[export]{adjustbox}

\usepackage{color}
\definecolor{myurlcolor}{rgb}{0,0,0.7}
\definecolor{myrefcolor}{rgb}{0.8,0,0}
\definecolor{colormult00}{RGB}{86, 180, 233}%
\definecolor{colormult01}{RGB}{230, 159, 0}%
\definecolor{colormult02}{RGB}{204, 121, 167}%
\definecolor{colormult03}{RGB}{240, 228, 66}%
\definecolor{colormult04}{RGB}{0, 158, 115}%
\definecolor{colormult05}{RGB}{213, 94, 0}%
\definecolor{colormult06}{RGB}{0, 114, 178}%
\definecolor{colormult07}{RGB}{0,0,0}
\definecolor{bronze}{rgb}{0.8, 0.5, 0.2}
\usepackage[unicode=true,pdfusetitle, bookmarks=false,bookmarksnumbered=false,
bookmarksopen=false, breaklinks=false,pdfborder={0 0 0},backref=false,
colorlinks=true, linkcolor=myrefcolor,citecolor=myurlcolor,urlcolor=myurlcolor]
{hyperref}
\DeclareUnicodeCharacter{2028}{\linebreak}

\definecolor{orange}{RGB}{255,165,0}
\definecolor{darkgreen}{RGB}{0, 170, 0}
\definecolor{pink}{RGB}{237,16,118}

\usepackage{lipsum}

\begin{document}
\title{Simulating quantum repeater strategies for multiple satellites}
 
 \author{Julius~Wallnöfer}
 \affiliation{Dahlem Center for Complex Quantum Systems, Freie Universit\"at Berlin, %
 14195 Berlin, Germany}
 
  \author{Frederik~Hahn}
 \affiliation{Dahlem Center for Complex Quantum Systems, Freie Universit\"at Berlin, %
 14195 Berlin, Germany}
 
 \author{Mustafa~Gündoğan}
\affiliation{Institut f\"{u}r Physik, Humboldt-Universit\"{a}t zu Berlin, %
Berlin 12489, Germany} 
 \affiliation{Deutsches Zentrum für Luft- und Raumfahrt e.V. (DLR), Institute of Optical Sensor Systems, 
 12489 Berlin, Germany}
 \author{Jasminder~S.~Sidhu}
\affiliation{SUPA Department of Physics, University of Strathclyde, John Anderson Building, %
Glasgow, G4 0NG, UK}

 \author{Fabian~Wiesner}
 \affiliation{Dahlem Center for Complex Quantum Systems, Freie Universit\"at Berlin, %
 14195 Berlin, Germany}

  \author{Nathan~Walk}
 \affiliation{Dahlem Center for Complex Quantum Systems, Freie Universit\"at Berlin, %
 14195 Berlin, Germany}
 
 \author{Jens~Eisert}
 \affiliation{Dahlem Center for Complex Quantum Systems, Freie Universit\"at Berlin, %
 14195 Berlin, Germany}
 \affiliation{Helmholtz-Zentrum Berlin für Materialien und Energie, %
 14109 Berlin, Germany}

 \author{Janik~Wolters}
 \affiliation{Deutsches Zentrum für Luft- und Raumfahrt e.V. (DLR), Institute of Optical Sensor Systems, 
 12489 Berlin, Germany}
  \affiliation{Technische Universität Berlin, Institut für Optik und Atomare Physik,
  10623 Berlin, Germany}
\date{\today}

\begin{abstract}
    A global quantum repeater network involving satellite-based links is likely to have advantages over fiber-based networks in terms of long-distance communication, since the photon losses in vacuum scale only polynomially with the distance -- compared to the exponential losses in optical fibers. To simulate the performance of such networks, we have introduced a scheme of large-scale event-based Monte Carlo simulation of quantum repeaters with multiple memories that can faithfully represent loss and imperfections in these memories. In this work, we identify the quantum key distribution rates achievable in various satellite and ground station geometries for feasible experimental parameters. The power and flexibility of the simulation toolbox allows us to explore various strategies and parameters, some of which only arise in these more complex, multi-satellite repeater scenarios. As a primary result, we conclude that key rates in the  kHz range are reasonably attainable for intercontinental quantum communication with three satellites, only one of which carries a quantum memory.
\end{abstract}
\maketitle

\section{Introduction}

Our modern networked societies are more dependent than ever on highly secure data transmission. Typical examples are
constituted by the control of critical infrastructures -- such as energy generation, communication, transportation and logistics -- as well as the exchange of health data. 
The digital encryption methods used today offer a range of attack points that can be overcome with the help of quantum key distribution (QKD). It is therefore desirable to establish QKD in a future multi-level digital security architecture in addition to the technology already in use.

A global quantum communication network with satellite-based links is likely to have advantages over fiber-based networks in terms of long-distance QKD, since the exponential photon losses introduced by optical fibers are too detrimental for distances beyond a few hundred km~\cite{Diamanti2016, Chen2020}. \emph{Quantum repeaters (QRs)}~\cite{Briegel1998, Sangouard2011} have been proposed to push this limit further. Here, intermediate, untrusted repeater stations involving distillation and swapping~\cite{Pan1998} steps reminiscent of 
quantum teleportation~\cite{TeleportationReview} allow the fundamental limitations of direct quantum communication to be overcome.
Although fiber-based QRs offer distances well beyond the direct communication limit, largely governed by the \emph{repeaterless  bound} (the PLOB bound) \cite{PLOB}
(see also Refs.~\cite{PhysRevLett.102.050503,Takeoka2014, Wilde2017}), 
and in principle allow for secure communication between arbitrary distances, they are still limited to around a few thousand km~\cite{Sangouard2011, Vinay2017} which precludes their use for global quantum networking. 

In contrast, 
\emph{satellite-based free-space QKD} (satQKD) benefits from a polynomial scaling with distance. In the field of satQKD, there are already multiple studies~\cite{Sidhu2020_npjQI, Lim2021_PRL, sidhu2021key} supporting numerous initiatives and missions both active and in planning~\cite{CSA_QEYSSat,Haber2018,Kerstel2018,Mazzarella2020, Villar2020} phases from Europe, North America and Asia relying on the BB84 and BBM92 schemes~\cite{Sidhu2021}. The most prominent example is the MICIUS which has realized many milestone demonstrations including teleportation from ground to satellite~\cite{Ren2017}, decoy-state BB84 QKD from satellite to ground~\cite{Liao2017} and a long distance, entanglement-based QKD with the BBM92 protocol~\cite{Yin2020}. The ranges in these experiments have been limited to the line-of-sight distance of the satellite which depends on its orbit. MICIUS has further demonstrated a beyond-line-of-sight QKD between Vienna and Beijing, operating as a trusted node~\cite{Liao2018}.

However, untrusted node operation for beyond-line-of-sight distances towards truly global scales requires the implementation of a QR protocol enabled by on-board \emph{quantum memories} (QMs)~\cite{Gundogan2021, Liorni2021}. Furthermore, QMs do not only help increase the overall network range but also offer a solution to low detection rates in entanglement-based schemes~\cite{Abruzzo2014,Panayi2014, Boone2015, Luong2016, Trenyi2020, Bhaskar2020, Langenfeld2021} and thus facilitate \emph{memory-assisted QKD} (MA-QKD) protocols which can be thought of as a single-node QR link. By synchronizing otherwise probabilistic detections, a single repeater station with a QM would change the scaling of the key rate from $\eta_{\text{ch}}$ to $\sqrt{\eta_{\text{ch}}}$ where $\eta_{\text{ch}}$ is the channel transmission. The development of systems for MA-QKD will not only enable the broad commercial use of satellite QKD, but will also promote the exploitation of other quantum technologies. Almost all key components for MA-QKD with untrusted satellites are already available or very well developed. This includes optical terminals, single photon sources, and detectors. Only quantum memories have a relative research backlog~\cite{Heshami2016}. These developments are forward looking and promising, but at the same time it is far from clear how to optimally devise schemes for satellite based quantum key distribution with realistic resources.

In this work, we comprehensively analyze multiple quantum repeater schemes that rely on satellites with quantum memories for continental and intercontinental distances~\cite{Rodriguez2021},  substantially going beyond our earlier work~\cite{Gundogan2021}. We make use of an event-based Monte Carlo simulation that enables the analysis of quantum repeaters with multi-mode memories. We simulate achievable MA-QKD rates in different satellite and ground station geometries for current and near-future experimental parameters. In addition, the current work utilizes memory cut-off times~\cite{Rozpedek2018} to improve achievable key rates, and 
stresses the importance of choosing them appropriately.

\section{Results}

\begin{figure*}
    \includegraphics[width=0.9\linewidth]{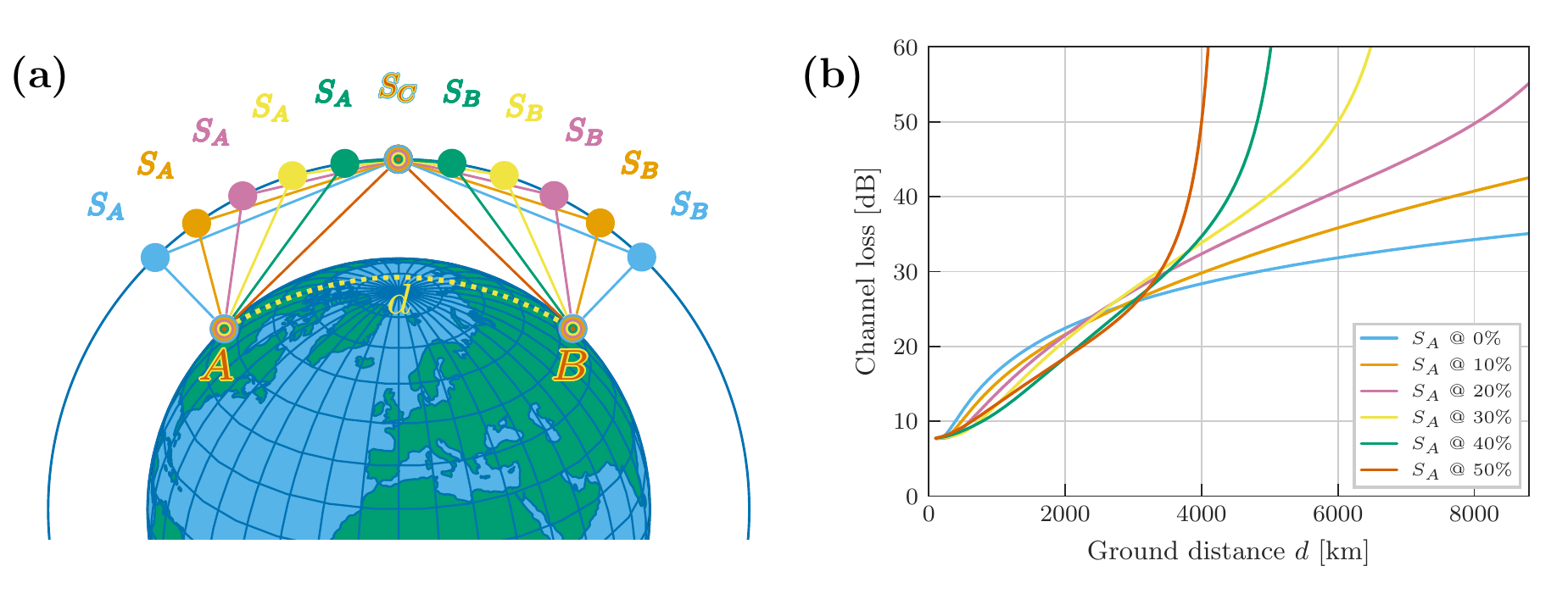}
    \caption{\label{fig:configuration_loss}(a) Using three satellites to reach distances beyond the horizon. The positioning of the satellites is a new parameter to optimize in this scenario. (b) Total loss when trying to establish an entangled link between $A$ and $S_C$ with an entangled pair source at $S_A$.  $S_A @ \mathrm{x}\%$ denotes that satellite $S_A$ is positioned vertically above the point at $x\%$ of the total ground distance. With three satellites one can avoid sending qubits through the atmosphere at a very shallow angle by positioning the outer satellites closer to the ground stations. Orbital height $h=400\mathrm{km}$, divergence angle $\theta_d=3\mu\mathrm{rad}$.}
\end{figure*}

For establishing quantum communication over intercontinental distances, it has been shown that making use of a satellite with a quantum memory can provide an advantage over systems without memory \cite{Gundogan2021, Liorni2021}, e.g., higher key rates for setups with only one satellite and shorter entanglement distribution times for multiple satellites. Quantum memories in principle suited for satellites have been demonstrated, e.g., in Ref.~\cite{Rodriguez2021}. In this work we build upon these earlier results and further analyze scenarios utilizing multiple satellites as repeater stations for a quantum repeater.

To this end, we have developed a large scale numerical Monte Carlo simulation for quantum repeaters that can faithfully represent loss and imperfect quantum memories, as well as other sources of noise such as dark counts. While there are existing approaches dealing with the computation of key rates for different repeater setups \cite{Sangouard2011, Luong2016, Trenyi2020, Shchukin2019}, the generalization of these methods to longer distances and other error models is by no means straightforward, e.g., an analytical approach also involves an intricate analysis of entanglement swapping strategies \cite{Shchukin2021}. Another challenge lies in the fact that for setups with multiple repeater links, a trial to establish an entangled pair somewhere along the line being successful or not can potentially influence the wait times in quantum memories everywhere in the setup. Therefore, we have chosen a simulation as our approach.

To be specific and close to the desiderata pertaining to realistic implementations, we focus on scenarios that make use of three satellites. When trying to reach very long distances, approaches with only one satellite will invariably reach a limit where the connection between ground stations becomes geometrically impossible or at least suffer from very high loss due to a shallow transmission angle through the atmosphere. While picking a higher orbit for the single satellite can extend the range, there is a significant trade-off in having to send photons much longer distances. For a proper comparison one needs to also take the different orbital period into account, which we touch upon in our analysis in Section \ref{sec:satellite_path}.

While the movement of the satellites is indeed essential (and will considered later), first, we consider the following setup with static satellite, which already contains a breadth of effects to analyze:
Two ground stations $A$ and $B$ are separated by ground distance $d$. Three satellites are used to establish a secret key between them. The central satellite $S_C$ is positioned halfway between the ground stations at orbital height $h$, however, the other two satellites $S_A$ and $S_B$ can be positioned at the same orbital height at any distance from the ground station and the central satellite as depicted in Fig.~\ref{fig:configuration_loss}a.

The positioning of the satellites becomes an additional decision for such a setup with three satellites, which is not present when using only one satellite. The two main sources of loss are the elevation angle dependent atmospheric loss $\eta_\mathrm{atm}(\theta)$ and the distance dependent diffraction loss $\eta_\mathrm{dif}$. Picking the position of satellites $S_A$ and $S_B$ clearly comes with a trade-off between those two sources of loss. Positioning the satellite directly above the ground station minimizes the atmospheric loss, but also means photons will need to be sent over longer distances. 

In Fig.~\ref{fig:configuration_loss}b the total loss for establishing a link between $A$ and $S_C$ with an entangled pair source at $S_A$ is shown for different positions of the satellites $S_A$ and $S_B$. There is a trade-off between avoiding as much atmospheric loss as possible when $S_A$ is positioned right above $A$ and the longer distance between the satellites that causes. However, for the very tightly collimated, diffraction-limited beams we consider here ($\theta_d = 3\mu\mathrm{rad}$) it becomes clear that for long distances it is advantageous to avoid as much atmospheric losses as possible. For a system with higher diffraction losses, this trade-off may not be as clear and would need to be reconsidered.
A detailed discussion about the error model and a description of the protocols can be found in the Appendices \ref{sec:error_models} and \ref{sec:protocols}, respectively.

\begin{figure*}
    \includegraphics[width=0.94\linewidth]{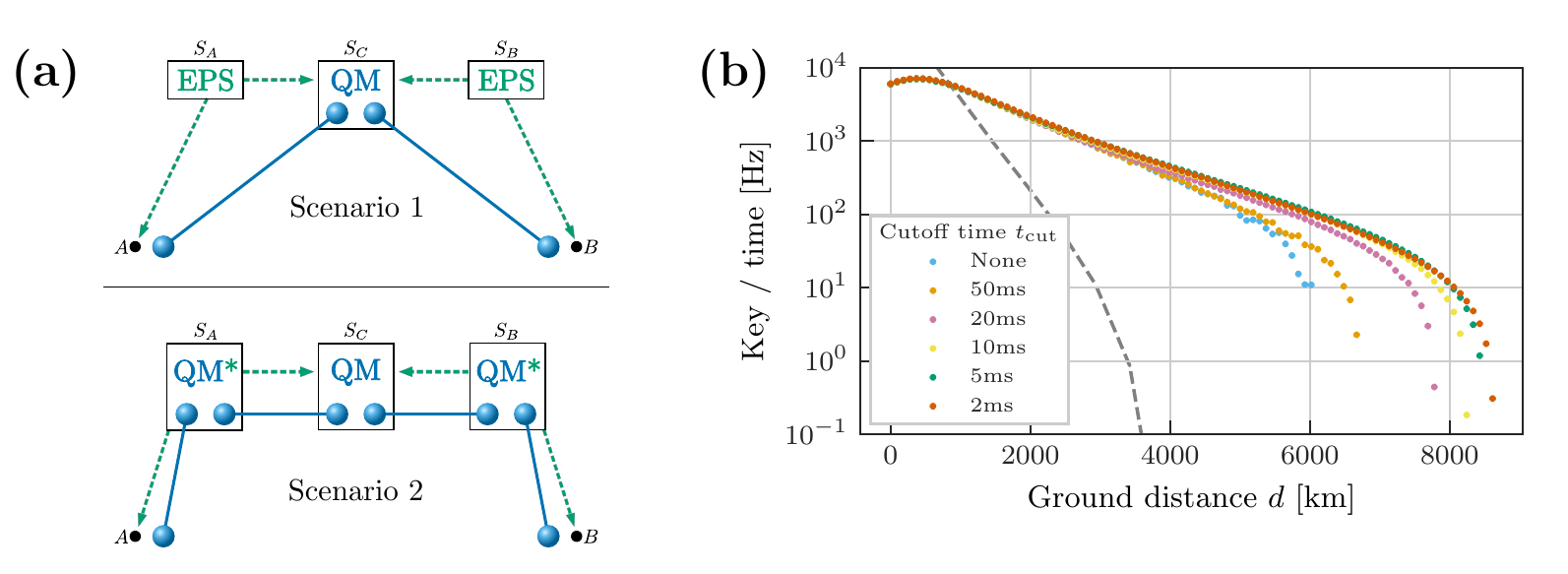}
    \caption{\label{fig:scenarios_cutoffs}(a) Illustration of the considered scenarios. Satellites $S_A$ and $S_B$ are equipped with entangled pair sources (EPS) and send qubits to the other satellite and the ground stations (dashed lines). In Scenario 1 only the central satellite $S_C$ has quantum memories (QM) and two repeater links are established. In Scenario 2 all of the satellites have quantum memories and a protocol with four repeater links are used. (b) Scenario 1: Choosing an appropriate cutoff time $t_\mathrm{cut}$ -- a maximum time a qubit is kept in a quantum memory -- can be very beneficial to the keyrate. The optimal value for $t_\mathrm{cut}$ is distance dependent and it is possible to choose a value that is too low.
    Dashed line indicates a BBM92 protocol with only one satellite and a clock rate of 20MHz with the same loss model.
    This plot is for $\theta_d = 6\mu\mathrm{rad}$ as the effect is more pronounced in situations with high loss.}
\end{figure*}

\subsection{Scenario 1}
 In this scenario, we look at an approach where the outer satellites establish links between the central satellite $S_C$ and each ground station and $S_C$ connects them by entanglement swapping, see Fig.~\ref{fig:configuration_loss}a. Hence, this is a protocol consisting of only two repeater links. For this setup only $S_C$ needs to be equipped with the ability to store qubits in a quantum memory, so we consider that the central satellite has two quantum memories with $n$ modes each as well as the ability to perform entanglement swapping. We assume that all quantum memories in both scenarios are directly heralding memories, i.e., it is immediately possible to tell whether loading a qubit into memory has been successful. $S_A$ and $S_B$ both are equipped with an entangled pair source that allows them to distribute entangled pairs between their assigned ground station and $S_C$.
 
 The basic idea of this protocol is very similar to one with only a single satellite, because only two repeater links are needed. Links are established between the central satellite and the ground stations and the central station has to wait for confirmation that a photon has arrived at each ground station before performing entanglement swapping. However, there are some subtle differences in terms of timing that need to be considered.
 
 The main difference from protocols in
 Refs.~\cite{Luong2016, Trenyi2020} is that the satellites with entangled pair sources do not have the information about whether a qubit has been loaded into memory successfully. Therefore, it makes sense for the source not to wait until confirmation from the satellite with the memory but to instead continuously send out entangled pairs. Therefore one central limitation lies in the maximum possible rate of entangled pair generation $f_\mathrm{clock}$ by the source.
 
 The continuous sending of pairs is conceptually similar to the up-link scenario considered in Ref.~\cite{Gundogan2021} with the difference that here the entangled pair sources are located on board of satellites instead of being on the ground. However, here the waiting times cannot be eliminated completely as the central satellite still has to wait for confirmation that a qubit arrived at the ground station before performing the entanglement swapping operation.

\subsection{Scenario 2}

In this alternative scheme we instead consider a setup that establishes four repeater links and uses three successive entanglement swapping operations to finally connect the ground stations $A$ and $B$. This means all three satellites need to have the capability to store qubits in a directly heralded fashion. We assume that similar to the above scenario each satellite is equipped with two quantum memories with $n$ modes each~\cite{Trenyi2020}. However, this time the satellites $S_A$ and $S_B$ contain emissive QMs (denoted by * in Fig.~\ref{fig:scenarios_cutoffs}a) that are able to emit single photons that are entangled with an internal atomic excitation, i.e., a stored matter qubit~\cite{Kutluer2019, Langenfeld2021}. Satellite $S_C$, on the other hand, carries absorptive type QMs~\cite{Gundogan2015, Wolters2017} that are capable of catching a flying qubit for storage similar to the QMs in Scenario 1.

The satellites that generate entangled states ($S_A$ and $S_B$) continuously try to establish new links with their neighbouring stations. This means that whenever a memory mode at $S_A$ is empty, a new entangled memory-photon pair is generated. The associated photonic qubit is then sent to the other station --- either the associated ground station $A$ or the central satellite $S_C$. The memory qubit at $S_A$ will need to be stored at least until confirmation from the other repeater station is received to confirm whether this trial has been successful. For this protocol, we assume that the trial for multiple memory modes can run in parallel and independently of each other. However, in practice, it is likely that instead of spatially separated channels there will be a number of time slots available in a shared channel. As long as the number of available time slots is much larger than the number of memory modes, the effects of sharing a channel are negligible. This is in contrast to Scenario 1, where the amount of photons that can be sent through the channel in a given time (which we directly link to the rate of entangled state generation for our simulation) is actually a limiting factor.

Furthermore, one needs to pick a strategy for how the entanglement swapping operations are handled if multiple successfully established qubits are sitting in memory, which is a situation that can potentially occur at each of the three satellites. For this scenario we chose to perform entanglement swapping as soon as it becomes available (i.e., there is no fixed order in which the satellites need to perform their Bell measurements) and always pick the eligible qubits that have sat in memory for the longest time. While this is likely not optimal (e.g., a more recently established entangled pair will have a higher fidelity), it completely eliminates the need for additional two-way communication (and therefore additional waiting times) between the satellites about the entanglement swapping process, that would arise from more involved strategies.

Compared to Scenario 1 this setup certainly introduces an operational overhead in the form of more qubits having to wait in quantum memories (and therefore dephase) as well as the need for additional operations. It is nonetheless interesting to consider as it is likely that this type of scheme will become more relevant once more advanced quantum repeater protocols, e.g., with added entanglement purification, become experimentally feasible.
 
\subsection{Key rates}

The asymptotic key rate is lower bounded by \cite{Luong2016, Lo2005, Scarani2009}
\begin{equation}
    r \left[1 - h(e_X) - f h(e_Z) \right],
\end{equation}
where $r$ is the rate of bits obtained from successful coincidence measurements that correspond to valid entanglement swapping operations, $h$ is the binary entropy function and $e_{X(Z)}$ is the quantum bit error rate in the $X$ ($Z$) basis.
We assume that the error correction inefficiency $f$ is equal to $1$. 
We let the simulation run until we have generated a large sample \footnote{$10^5$ for Scenario 1 and $10^4$ for Scenario 2.} of long-distance links between $A$ and $B$, which we use to calculate the sample mean for $r$ and $e_{X(Z)}$.

Ultimately, a finite-size, composably secure analysis is essential for cryptographic applications \cite{Tomamichel2012}. Moreover, in a practical setting the effects of finite data blocks on satellite quantum key distribution  (see, e.g.,
Refs.~\cite{Bacco2013,Lim2021_PRL}) can potentially be particularly significant if achievable block sizes in a single pass are limited. Nevertheless, the asymptotic rates are still informative since they provide an upper bound to the performance limit of satellite-based quantum repeater strategies and are reflective of performance for reasonable block sizes. 

\subsection{Cutoff times}
In essence, using quantum memories for a quantum repeater allows one to trade some of the probability that measurements at the ground stations coincide for an overall higher rate of qubits that successfully arrived at their destination --- with the entanglement swapping operation allowing one to connect two links that are more likely to be successful individually.

However, if qubits are stored in quantum memories for too long, the additional dephasing at some point becomes too detrimental and can reduce the achievable key rate. As an optimization, it is therefore important to add a mechanism to discard qubits that have dephased too much. One simple mechanism is to choose a cutoff time $t_\mathrm{cut}>0$, which is the maximum time a qubit is allowed to sit in memory \textit{after} a successful generation of an entangled link is confirmed. Such a cutoff mechanism has been previously proposed, e.g, in Ref.~\cite{Rozpedek2018}, and its inclusion is one of the primary improvements of the protocol compared to the previous results in Ref.~\cite{Gundogan2021}. Discarding leftover qubits in memory after each cycle in the protocol in Ref.~\cite{Trenyi2020} also has a similar effect to prevent too much dephasing noise from building up on the qubits in memory.

By carefully tuning the cutoff time, one can essentially choose how much of the trade-off mentioned above is acceptable. However, it should be noted that this is not the only possible mechanism for choosing when to discard dephased qubits. In fact, finding the optimal strategy at every time step has been shown to 
require resources that are scaling exponentially in the input size
\cite{Khatri2021policies}.

In Fig.~\ref{fig:scenarios_cutoffs}b the effect different choices of $t_\mathrm{cut}$ can have on achievable distances and key rates is demonstrated. While the precise impact depends on many factors, e.g., the relation of loss rates and memory quality, choosing an appropriate cutoff time is crucial to extend the reachable distance. It should be noted that choosing a too short cutoff time can actually be detrimental to the key rate, e.g., in Fig.~\ref{fig:configuration_loss}b the achievable key rate is higher for $t_\mathrm{cut} = 5\text{ms}$ than for $2\text{ms}$ in the $5000$-$7000 \text{km}$ range.
This effect can be easily understood in the most extreme case as a very low $t_\mathrm{cut}$ will essentially turn the protocol into a quantum repeater without memories.

\subsection{Achievable key rates for realistic parameters}
Having a numerical simulation opens up the possibility to investigate a large range of values for all relevant error parameters. However, in order to interpret the results it is important to choose a meaningful parameter set. In Table \ref{tab:base_params} we list the parameters for our error model, which are considered the baseline for our simulation. These are chosen according to realistic ranges for current or near-term implementations. 

\begin{table}[htbp]
    \centering
    \includegraphics[width=\linewidth]{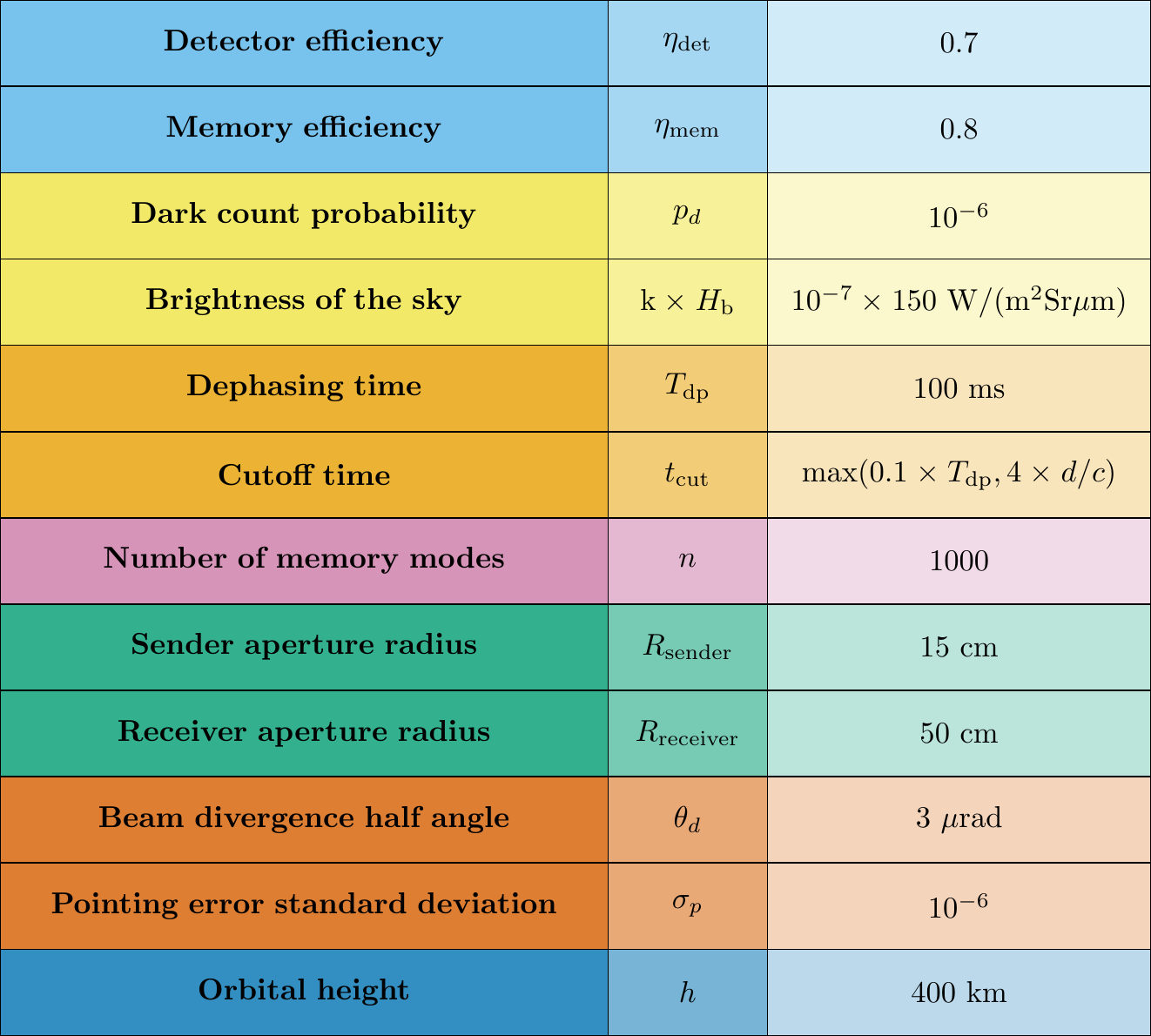}
    \caption{The simulation allows us to explore a range of parameters. These are the base parameters for our simulation that correspond to realistic ranges for current or near-term implementations. All deviations from this set for certain scenarios are highlighted in the text.}
    \label{tab:base_params}
\end{table}

\begin{figure*}
    \includegraphics[width=0.9\linewidth]{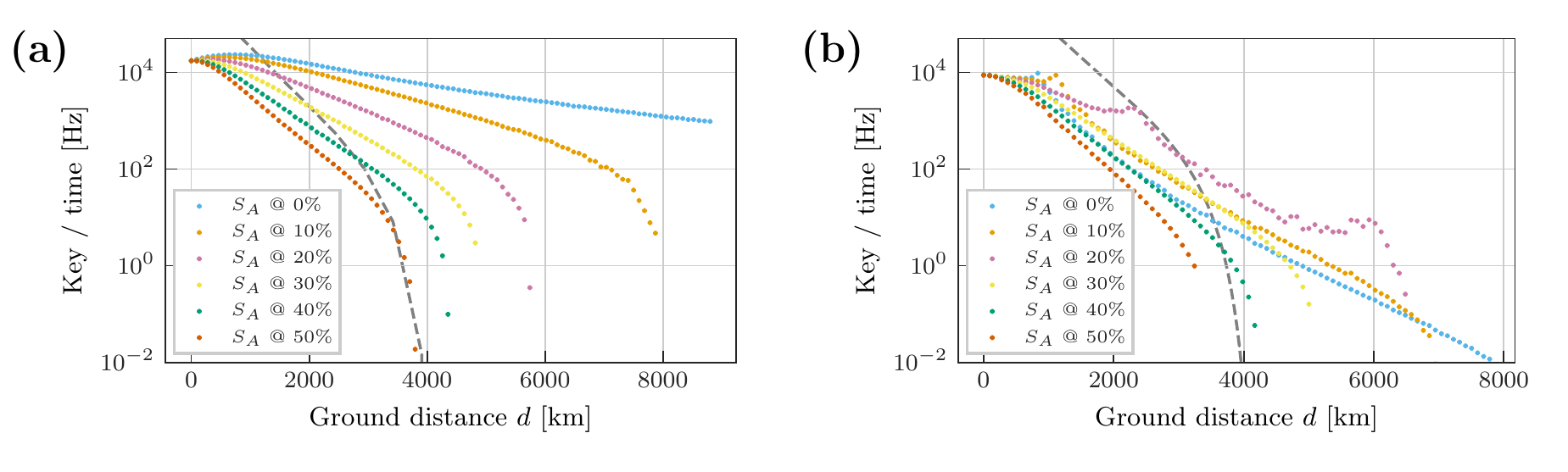}
    \caption{\label{fig:sat_positions}Achievable key rates for different choices of positions of the satellite. This is a new parameter to optimize when using multiple satellites. Satellite $S_A$ is positioned vertically above varying percentages of the total ground distance. Dashed line indicates the BBM92 protocol with only one satellite and a clock rate of $20$~MHz. (a) Scenario 1, a protocol with outer satellites distributing one pair between the ground station and the central satellite each. Here positioning the satellites $S_A$ and $S_B$ directly above the ground station to avoid as much atmospheric noise as possible proves beneficial. (b) Scenario 2, a protocol that establishes four links of entangled pairs between satellites and stations. Here, the distance dependent trade-offs are more complex.}
\end{figure*}

As mentioned before, the position of the satellites $S_A$ and $S_B$ is a new decision that has to be made when using more than one satellite. In 
Fig.~\ref{fig:sat_positions} the key rate of multiple positions is shown for the base parameter set in Table \ref{tab:base_params}. For this parameter set it is obvious that avoiding as much atmospheric loss as possible is worth the additional diffraction loss from the longer distance between $S_A$ and $S_C$ in Scenario 1, even for short communication distances. However, for the more involved Scenario 2 consisting of four repeater links, positioning $S_A$ directly above ground station $A$ is not optimal. Hence, both the loss parameters and the precise choice of protocol influence the optimal satellite positions. 

One interesting thing to note is that for some configurations the key rate does not strictly decrease with the distance. This is due to a situation that can happen with multi-mode memories, if multiple qubits sit in memory waiting for the other side to be ready for entanglement swapping. When the loss is much smaller for one repeater segment than the other (as is the case for the very asymmetric losses in Scenario 2), this is something that will happen often, even with when using a cutoff time strategy. While most of the established pairs will not be swapped due to the the high-loss segment not providing pairs fast enough, the average time a pair that does end up getting swapped is sitting in memory can end up being lower if both segments have comparable losses. Therefore, unintiutively an increase in the loss of the comparatively low-loss segment can lead on average to a higher fidelity of swapped entangled pairs. Indeed, the local maxima for the case we consider here are found where the (distance-dependent) loss for the channel between $A$ and $S_A$ is comparable to the inter-satellite loss between $S_A$ and $S_C$. The keyrates could likely be improved further by optimizing $t_\text{cut}$ at each data point, or modifying the swapping strategy when multiple qubits are waiting in memory.

\begin{figure*}
    \includegraphics[width=\linewidth]{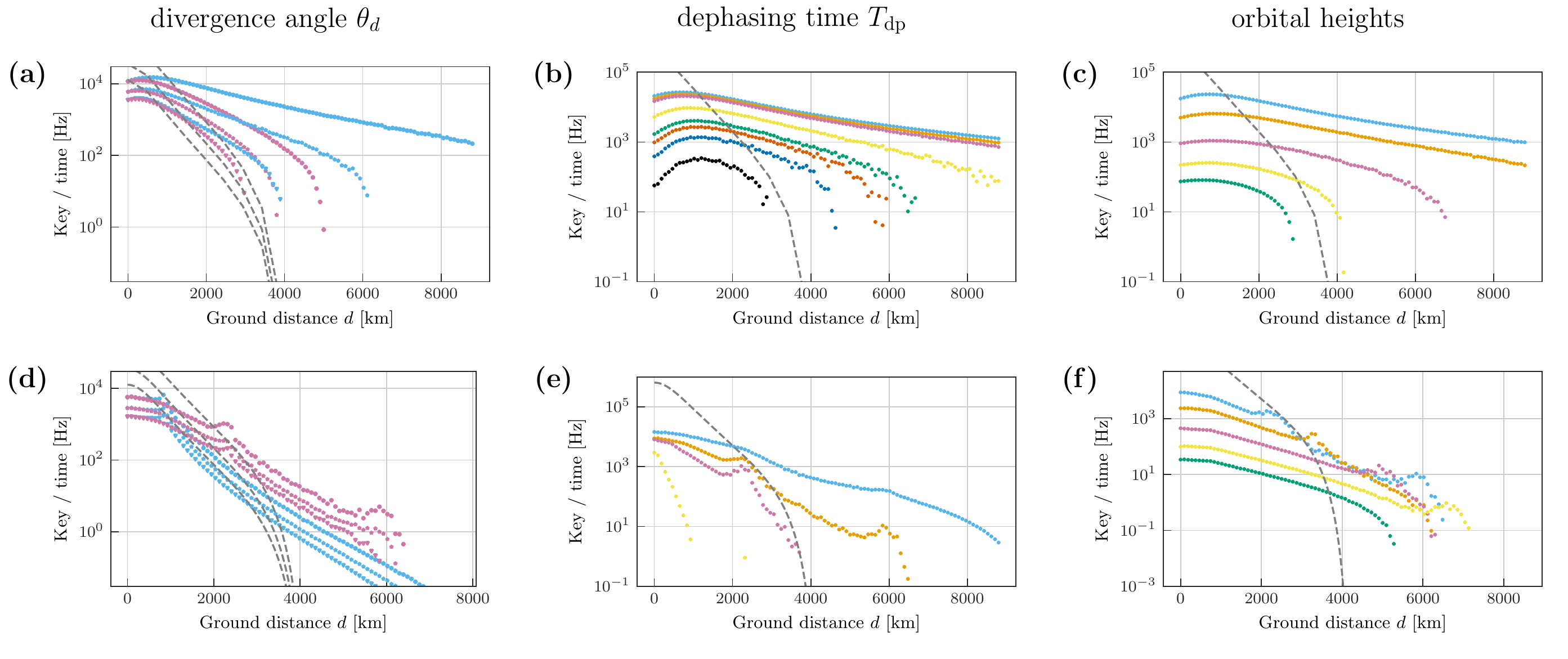}
    \caption{\label{fig:params_various}Exploring variations of the parameters. In each subfigure one parameter is varied, while the others are kept at their base value in Table \ref{tab:base_params}. All plots are made for 1000-mode quantum memories. a-c) show Scenario 1 (two repeater links), d-f) show Scenario 2 (four repeater links). a/d) Higher divergence angle $\theta_d = 4\mu\mathrm{rad}, 6\mu\mathrm{rad}, 8\mu\mathrm{rad}$ and therefore higher loss for $S_A$ satellite positions $0$ (blue) and $0.2$ (purple). b/e) Various memory qualities with dephasing times $T_\mathrm{dp}$ of  $1\mathrm{s}$ (blue), $100\mathrm{ms}$ (orange), $50\mathrm{ms}$ (pink), $10\mathrm{ms}$ (yellow), $5\mathrm{ms}$ (green), $4\mathrm{ms}$ (red), $3\mathrm{ms}$ (dark blue), $2\mathrm{ms}$ (black). c/f) Differing orbital heights for all three satellites $400\mathrm{km}$ (blue), $600\mathrm{km}$ (orange), $1000\mathrm{km}$ (pink), $1500\mathrm{km}$ (yellow), $2000\mathrm{km}$ (green).}
\end{figure*}

In Fig.~\ref{fig:params_various} we explore the effect of varying some parameters of interest, namely the divergence angle $\theta_d$ of the beams connection ground stations and satellites, the quality of the quantum memories and the orbital height of the satellites. Figures \ref{fig:params_various}a and \ref{fig:params_various}d show that naturally, a higher $\theta_d$ and therefore higher loss impacts the key rate significantly. Despite the lower rates when loss is small, Scenario 2 actually proves more resilient against higher loss rates. Figs.~\ref{fig:params_various}b and \ref{fig:params_various}e clearly demonstrate that memory quality plays an important factor when determining reachable distances. While having satellites in higher orbits could be used to extend the reachable ranges even further, figures \ref{fig:params_various}c and \ref{fig:params_various}f demonstrate that there is a significant drop in key rates even for small $\theta_d$.

\subsection{Effective rates for orbiting satellites}
\label{sec:satellite_path}

In the previous sections we did not consider the movement of the satellites in an orbit around the Earth. While these static scenarios already show a wide range of effects and allow to draw conclusions about the importance of various parameters, the actual numbers obtained for the key rates would be more appropriate for far-future implementations with a large number of available satellites, which makes it likely to find sets of satellites close to optimal positions for large time periods. However, when analyzing near-term experiments with one or three satellites, the available time windows and changes of, e.g., diffraction and atmospheric losses along the path of the satellite become a vital component.

\begin{figure*}
 \includegraphics[width=\linewidth]{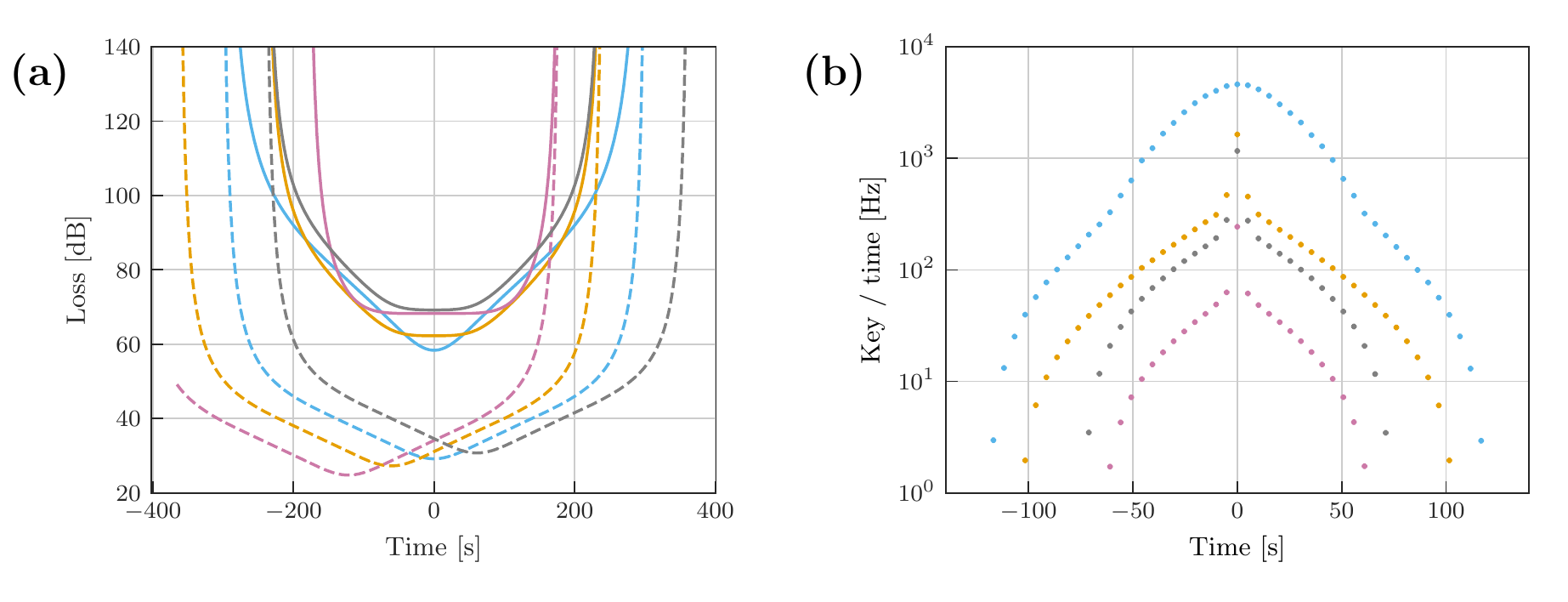}
 \caption{\label{fig:satellite_path} Three satellites passing above ground stations $d=4400\mathrm{km}$ apart. The colors match the relative positions of the satellites in Fig.  \ref{fig:sat_positions},
 such that when $S_C$ is exactly in the middle point between the ground stations $S_A$ will be at $0\%$ (blue), $10\%$ (orange), $20\%$ (pink) or $-10\%$ (gray) of the total ground distance $d$. (a) Combined atmospheric and diffraction loss between $A$ and $S_C$ (dashed lines) and along the whole optical path (solid lines). (b) The obtainable asymptotic key rate for Scenario 1 at points along the orbit if satellites were static at these positions.}
\end{figure*}

In the following we look at a fixed ground distance $d=4400\mathrm{km}$ (i.e., right at the edge where a single satellite at the same orbital height $h=400\mathrm{km}$ can no longer see both ground stations at the same time).
In Fig. \ref{fig:satellite_path}a, the total loss for establishing an entangled link between $A$ and $S_C$ with an entangled pair source at $S_A$ as the setup of three satellites travels along its orbit is shown. This means that depending on the spacing between satellites on the orbit, there are time windows of about $\sim$ 4-8 minutes where a signal can reach both ground stations. For comparison, the orbital period is approximately $92.4$ minutes for this orbital height.

While the asymptotic key rates one would obtain with static satellites at various points along the orbit (as shown in Fig. \ref{fig:satellite_path}b) are useful to get a sense of the performance of different setups, in order to estimate the actually obtainable rates one needs to analyse the effective quantum bit error rate of the raw bit strings collected at the ground stations. For each of the data points we calculate the quantum bit error rate and average it weighted by the raw bit rate. We perform this for both scenarios with three satellites and also compare them to protocols with one satellite at higher orbits, as an alternative way to make key distribution at this distance possible. The obtainable raw bits per pass as well as the effective key rate (taking into account the time waiting for the satellites to come in range again) are summarized for a selection of satellite configurations in Table \ref{tab:satellite_path_results}. For details of the calculations and additional results see Appendices \ref{sec:estimate} and \ref{sec:bright_night}.
It is worth noting that for the best working configurations around $10^4$-$10^6$ raw bits per pass of the satellites can be expected, which would be sensible block sizes for finite key distribution protocols \cite{Lim2021_PRL} without having to accumulate bits over multiple passes.

\begin{table*}[htbp]
    \centering
    \includegraphics[width=0.6\linewidth]{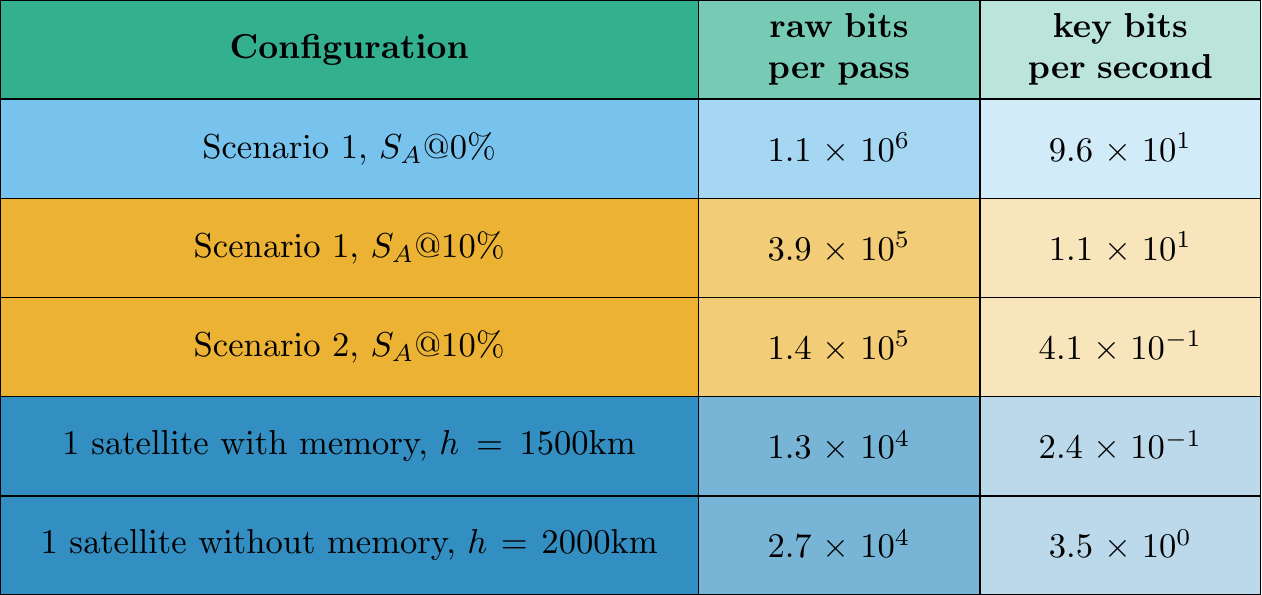}
    \caption{Obtainable bits per pass of the satellite configurations over the ground station as well as the effective key rate averaged over a whole orbital period for a selection of setups. Scenarios 1 and 2 describe different protocols using three satellites at orbital height $h=400\mathrm{km}$.}
    \label{tab:satellite_path_results}
\end{table*}

\section{Discussion}
In summary, we have developed a simulation for quantum repeaters that can deal with a variety of error models as well as multi-mode memories and setups with realistic protocols. This allows us to investigate scenarios that go beyond two repeater links (e.g., the protocol in Scenario 2 that uses four repeater links), for which no complete analytical description is known. We used this simulation to analyze the performance of schemes that use multiple satellites with quantum memories to perform QKD over long distances. Using more than one satellite allows one to reach distances that would be geometrically impossible with just one satellite. We have shown that reaching intercontinental distances with currently available or near-future experimental parameters is entirely feasible and even performing advanced schemes is reasonable, although the overhead of using multiple satellites is still very significant.
In the future we plan to use our simulation to analyze setups with actual experimental parameters to gauge the real performance that can be expected. Another direction would be to extend our approach to fully capture the effect of of changing conditions, e.g., 
as would be the case for analysing the full dynamics of moving satellites.

\section{Methods}
The main technique used to obtain our results is 
building on a substantial method involving
a real-time Monte Carlo simulation. Our method focuses on high-level decision making in creating protocols while faithfully including experimental parameters for many different physical implementations. This is in contrast to other large-scale simulations that put a much stronger emphasis on the network character of quantum networks, as being pursued, e.g.,
in Refs.~\cite{NetSquid, QuNetSim, QuISP}. We will report on substantial details and further applications of the simulation elsewhere. However, in the following, we briefly describe its basic working principles. 

\begin{figure}
    \centering
    \includegraphics[width=\linewidth]{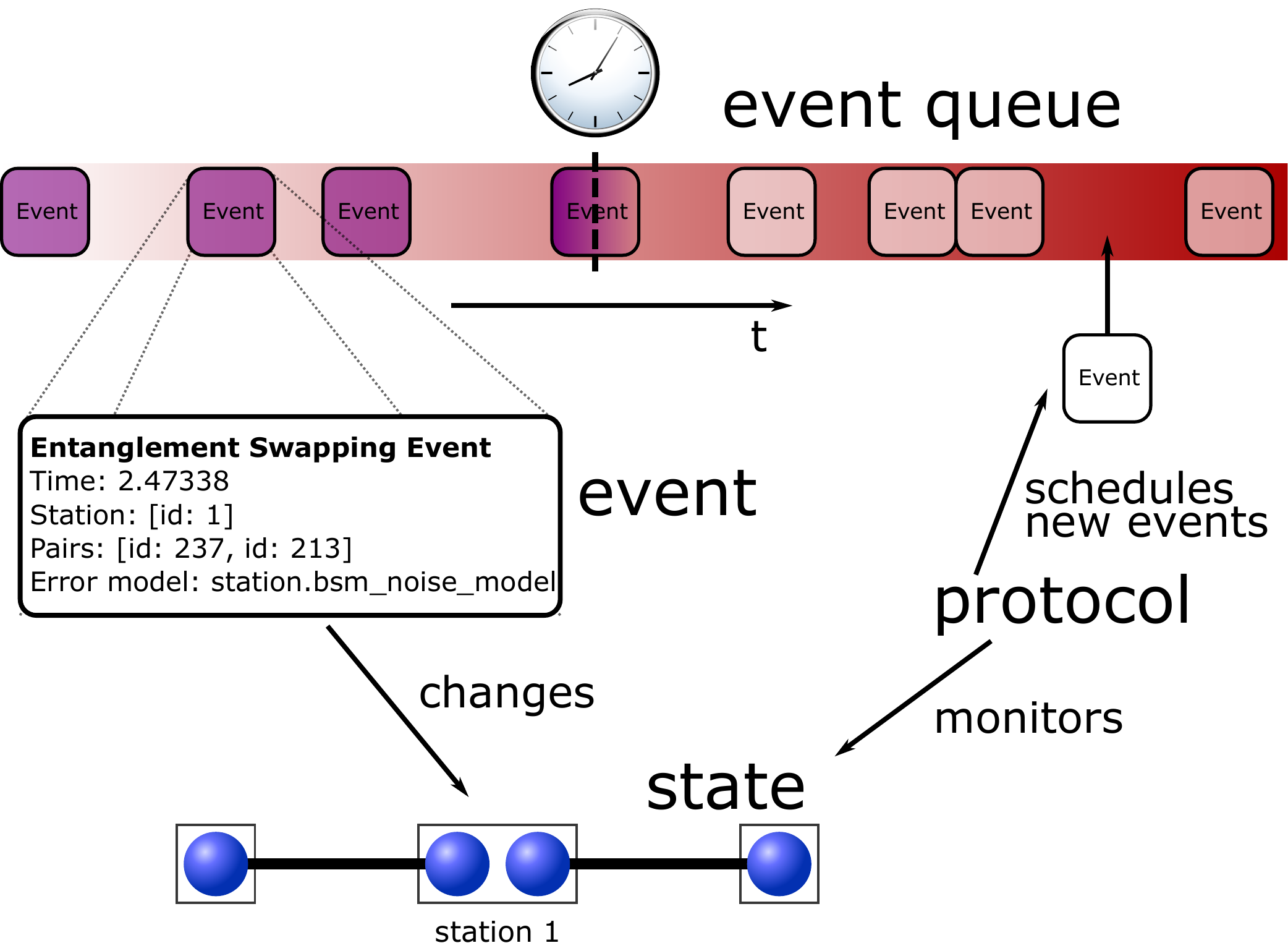}
    \caption{Schematic representation of the simulation framework. Events are scheduled in an event queue and resolved in order. The state of the simulation, i.e., the location of qubits and quantum state of the entangled pairs, is changed when events get resolved. The figure shows as an example some of the information associated with an entanglement swapping event: at which station the entanglement swapping operation is performed, which entangled pairs are involved in the process and which error model is relevant to the operation. The current state is monitored and new events are scheduled by the chosen protocol.}
    \label{fig:simulation_scheme}
\end{figure}

The simulation keeps track of the current situation, e.g., which pairs are currently established, at which stations the associated qubits are located and what the density matrix of each entangled pair is. 
All changes to the current situation happen via events that are scheduled in an event queue and resolved in order. For example, an event might be an entanglement swapping operation to connect two distant stations, or discarding a qubit that has been stored longer than the memory policy allows.
Furthermore, a protocol determines the strategy of what events are scheduled. For instance, one component of the protocol might consist of scheduling an event that generates a new pair if the quantum memory is empty and generating a new pair is not already scheduled.
An illustration of this scheme is depicted in 
Fig.~\ref{fig:simulation_scheme}.

We make use of two key methods that allow us to perform this simulation in a reasonable time frame. For one, we do not track individual photons that are much more likely to get lost than arrive at their destination when loss is high. Instead, we use the known success probability of distributing a pair in one trial $\eta$ and draw from a geometric probability distribution to determine how many sequential trials had to be performed in order to successfully establish one pair. This sampling from a probability distribution is why we call it a Monte Carlo simulation, even though other probabilistic aspects, e.g., dephasing noise in quantum memories and dark counts, are handled via the density matrix formalism. 
Secondly, we do not continuously update the effect of time-dependent dephasing noise in quantum memories, instead we only update the quantum state when it becomes relevant, which is possible because we keep track of when it was last updated. This ensures that having many events happen on other parts of the simulation does not cause an undue amount of calculation for unaffected parts.

\section{Data availability}
The raw output data of the simulation is available upon reasonable request.

\section{Code availability}
The source code that has been used to generate all results in this work is archived at DOI:~\href{https://doi.org/10.5281/zenodo.5603047}{10.5281/zenodo.5603047}.\\

\section{References}
\bibliography{references}

\section{Acknowledgements}
J.~Wa.~and J.~E.~acknowledge support from the BMBF (Q.Link.X and QR.X) and the Einstein Research Unit on Quantum Devices. F.~H.~acknowledges financial support from the German Academic Scholarship Foundation. N.~W.~has been funded by the European Union's Horizon 2020 research and innovation programme under the Marie Skłodowska-Curie Grant  Agreement No.~750905 and the DFG priority
program “Compressed Sensing in Information Processing - Phase 2 (CoSIP2). J.~Wo.~and M.~G.~acknowledge support from BMWi through DLR (QuMSeC, No.~50RP2090). M.~G. further acknowledges funding from the European Union's Horizon 2020 research and innovation programme under the Marie Skłodowska-Curie Grant Agreement 
No.~894590 and the support from DLR through funds provided by BMWi (OPTIMO, No.~50WM1958 and OPTIMO-II, No.~50WM2055). J.~S.~S.~acknowledge the support of EPSRC via the Quantum Communications Hub through grant number EP/T001011/1. We thank the HPC system of the Freie Universität Berlin \cite{hpc_fub} for computing time. 

\section{Author contributions}
J.~Wa.~wrote the code for the simulation framework. F.~H.~and F.~W.~contributed to analyzing and verifying the capabilities of the simulation framework. J.~Wa.~and F.~H. wrote the code for the protocols and obtained the numerical results presented in this work. 
M.~G., J.~S.~S.~and J.~Wo.~formulated the practically relevant loss models, noise models and provided experimental parameters. All authors contributed to picking relevant scenarios, developing appropriate protocols and writing the manuscript.

\section{Competing interests}
The authors declare no competing interests.

\appendix

\section{Error models}
\label{sec:error_models}
\subsection{Probability of establishing a link}
We consider several sources of imperfections that can cause an attempt to create an entangled pair between two stations to fail. The most prominent of these is photon loss, i.e.~the photon that is lost on its way between the stations.

For the diffraction loss we consider a model of paraxial Gaussian beam propagation.
The intensity profile at the sender is given by
\begin{equation}
    I(r, 0) = I_0 e^{-2 \frac{r^2}{w_0^2}}.
\end{equation}
$r$ is describing the distance from the optical axis and $w_0$ is the initial beam waist. $I_0$ is a constant that relates to the total power transmitted by the beam $P_0 = \frac{I_0 w_0^2}{2 \pi}$.
For diffraction limited beams there is a direct relation between how narrowly the beam can be collimated, the divergence half-angle $\theta_d$ and the wavelength $\lambda$:
\begin{equation}
 w_0 = \frac{\lambda}{\theta_d \pi}.
\end{equation}
As the beam travels along its path the beam widens according to
\begin{align}
    w(z) &= w_0 \sqrt{1 + \left(\frac{\theta_d}{w_0}\right)^2 z^2} \\
    I(r, z) &= I_0 \left(\frac{w_0}{w(z)}\right)^2 e^{-2 \frac{r^2}{w(z)^2}}.
\end{align}
At this point we also include another source of imperfection in the form of pointing error, i.e.,
the beam is not perfectly pointed at the target. Our model for the pointing error is very similar to the one described in Ref.~\cite{Bourgoin_2013}, but applied to the Gaussian beam widening model described above instead of using the more involved Rayleigh-Sommerfeld diffraction.

We assume the pointing error is randomly distributed according to a two-dimensional Gaussian distribution with standard deviations $\sigma_p$
\begin{equation}
    \frac{1}{2 \pi \sigma_p} e^{- \frac{r^2}{2 \sigma^2_p}},
\end{equation}
which means at distance $z$ the center of the intensity distribution is randomly shifted according to (for small pointing errors)
\begin{equation}
    g_z(r) = \frac{1}{2 \pi z \sigma_p} e^{- \frac{r^2}{2 (z \sigma_p)^2}}
\end{equation}
We express the resulting intensity distribution via the two-dimensional convolution
\begin{equation}
    \widetilde{I}(r,z) = \int_{0}^{\infty} \mathrm{d}r^\prime \int_{0}^{2\pi} \mathrm{d}\varphi \; r^\prime I(r^\prime, z) g_z(r^\prime).
\end{equation}
The power $P(z)$ that arrives at the receiver with with aperture radius $R_\mathrm{receiver}$ can then be calculated by integrating over the aperture
\begin{equation}
    P(z) = \int_{0}^{R_\mathrm{receiver}} \mathrm{d}r^\prime \int_{0}^{2\pi} \mathrm{d}\varphi \; r^\prime \widetilde{I}(r^\prime, z)
\end{equation}
Therefore, the probability of a photon arriving at a receiving station at distance $l$ is given by
\begin{equation}
    \eta_\mathrm{dif}(l) = \frac{P(l)}{P_0}.
\end{equation}
The effect of the receiver station potentially also not pointing in the correct distance is negligible for the cases we consider ($l \gg R_\mathrm{receiver}$, $\sigma_p \ll 1$) as long as the field of view of the receiver telescope is significantly larger than $\theta_d$ and $\sigma_p$.

\begin{figure*}
    \centering
    \includegraphics[width=0.8\linewidth]{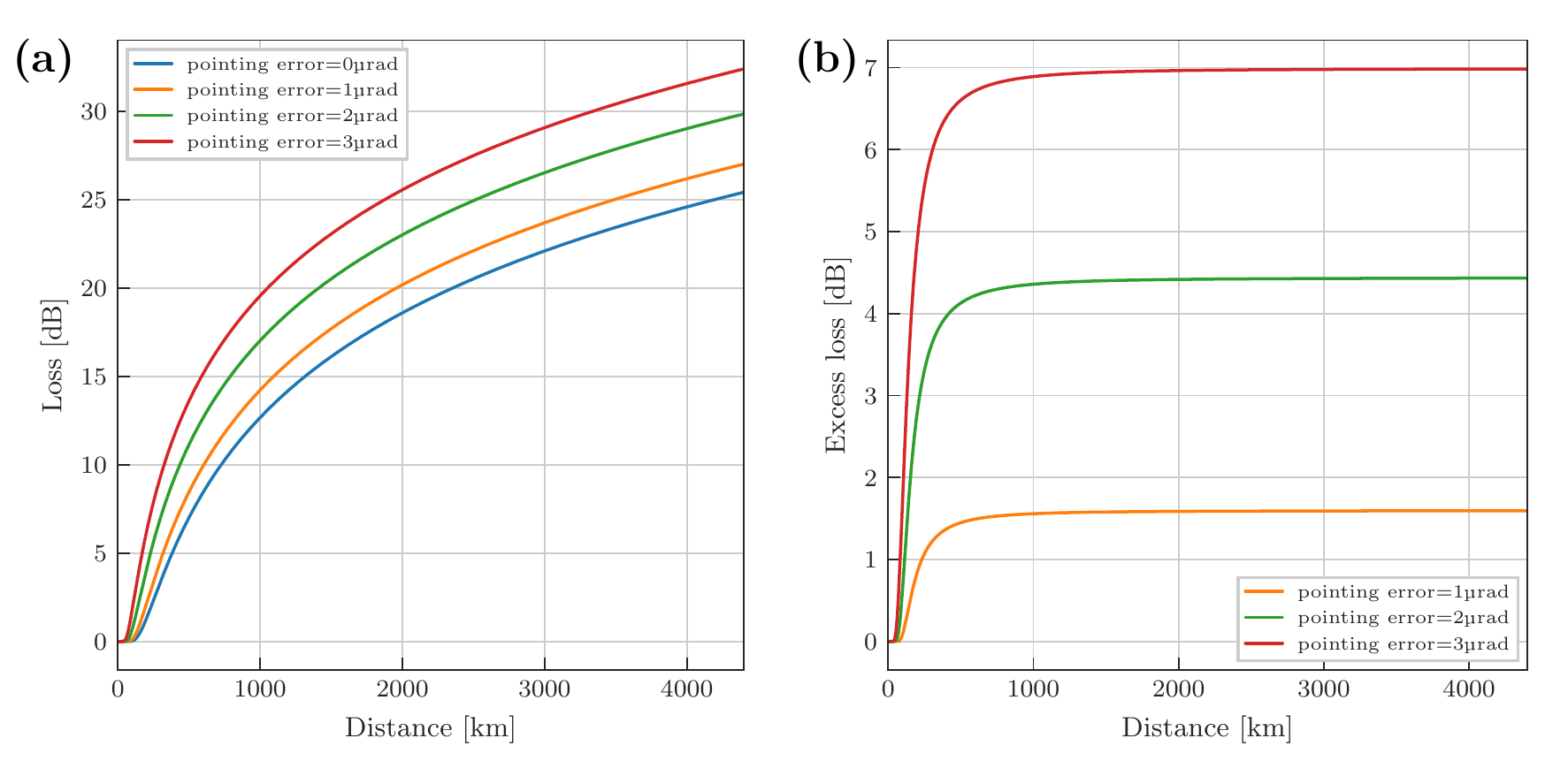}
    \caption{Effect of pointing error with standard deviation $\sigma_p$ depending on the distance. (a) Diffraction loss including pointing error $1 / \eta_\mathrm{atm}$ (b) Excess loss caused by the pointing error.}
    \label{fig:pointing_error}
\end{figure*}

In Fig.~\ref{fig:pointing_error} the effect of various values of $\sigma_p$ on the overall loss is shown. One can see that for long enough distances the additional loss stemming from pointing errors becomes almost constant.

When linking stations on the ground with satellites, the influence of the atmosphere must also be considered. The transmittivity of the atmosphere $\eta_\mathrm{atm}$ can be modeled as a function of the elevation angle $\theta$~\cite{Khatri2021}, as
\begin{equation}
    \eta_\mathrm{atm}(\theta) = \left(\eta_\mathrm{atm}^{\pi/2}\right)^{1/\sin\theta},
\end{equation}
where $\eta_\mathrm{atm}^{\pi/2}$ is the Zenith
transmittivity. We use $\eta_\mathrm{atm}^{\pi/2} = 0.8$ which is a reasonable value for a wavelength of $780\text{nm}$~\cite{Gundogan2021}.

Furthermore, establishing an entangled pair might also fail due to imperfections of the repeater stations. For the ground stations we use a detector efficiency $\eta_\mathrm{det} = 0.7$ and for the stations using quantum memories we use a memory efficiency of $\eta_\mathrm{mem} = 0.8$~\cite{Yang2016, Wang2019}.

The probability $\eta$ of successfully establishing an entangled pair is the product of the relevant efficiencies for a given link. For example, for a link between a ground station and a satellite hosting an emissive quantum memory capable of generating entangled states, the probability is given by
\begin{equation}
    \eta = \eta_\mathrm{mem} \eta_\mathrm{det} \eta_\mathrm{atm}(\theta) \eta_\mathrm{dif}(l).
\end{equation}
If the ground distance to the satellite is given by $L_g$ and the orbit altitude $h$, one can calculate the distance $l$ and the elevation angle $\theta$ by geometric considerations
\begin{align}
    l^2 &= {R_E^2 + (R_E + h)^2 - 2 R_E (R_E +h) \cos \left(\frac{L_g}{R_E} \right)},\\
    \theta &= \frac{\pi}{2} - \frac{L_g}{R_E} - \arcsin \left[\frac{R_E}{l} \sin\left(\frac{L_g}{R_E} \right) \right],
\end{align}
where $R_E = 6371 \text{ km}$ is the average earth radius.
In the scenarios that we consider in this work we have three types of lossy connections: ground-satellite, satellite-satellite and ground-satellite-satellite (the middle satellite uses its entangled pair source to connect a ground station and another satellite).

\subsection{Dark counts}
The detectors at the end stations naturally have to deal with dark counts.  With a dark count probability of $p_d$, the chance that the detector detects a click is naturally higher, resulting in an effective $\eta$
given by
\begin{equation}
    \eta_\mathrm{eff} = 1- ( 1 - \eta) (1- p_d)^2.
\end{equation}
The probability that a click is a real event is
determined as
\begin{equation}
    \alpha (\eta) = \frac{\eta (1-p_d)}{\eta_\mathrm{eff}}.
\end{equation}
This  dependency on
$\eta$
implies that when success rates are very low, most clicks are eventually caused by dark counts.
We model dark counts as local white noise, i.e.~if a click is caused by a dark count, we assume that the state for the putative photon that might have arrived is instead a fully mixed state. The inclusion of dark counts is modelled by a perfect detector preceded by a noisy channel given by
\begin{widetext}
\begin{equation}
    \mathcal{D}^{(i)}_w(\alpha) \rho = \alpha \rho + \frac{1-\alpha}{4} \left( \rho + X^{(i)} \rho X^{(i)} + Y^{(i)} \rho Y^{(i)} + Z^{(i)} \rho Z^{(i)} \right),
\end{equation}
\end{widetext}
where $i$ specifies that the channel acts on the $i$-th qubit.
The dark count rate is a characteristic of the detector -- usually given in dark counts per second-- whereas $p_d$ is the probability that a dark count occurs during a detection window that would lead to mistaking it for a real event. In this work we assume $p_d = 10^{-6}$, which corresponds to a dark count rate of a few Hz and detection windows of $\sim 1 \mathrm{\mu s}$.

\subsection{Background light}
\label{sec:background_light}

\begin{figure}
    \includegraphics[width=\linewidth]{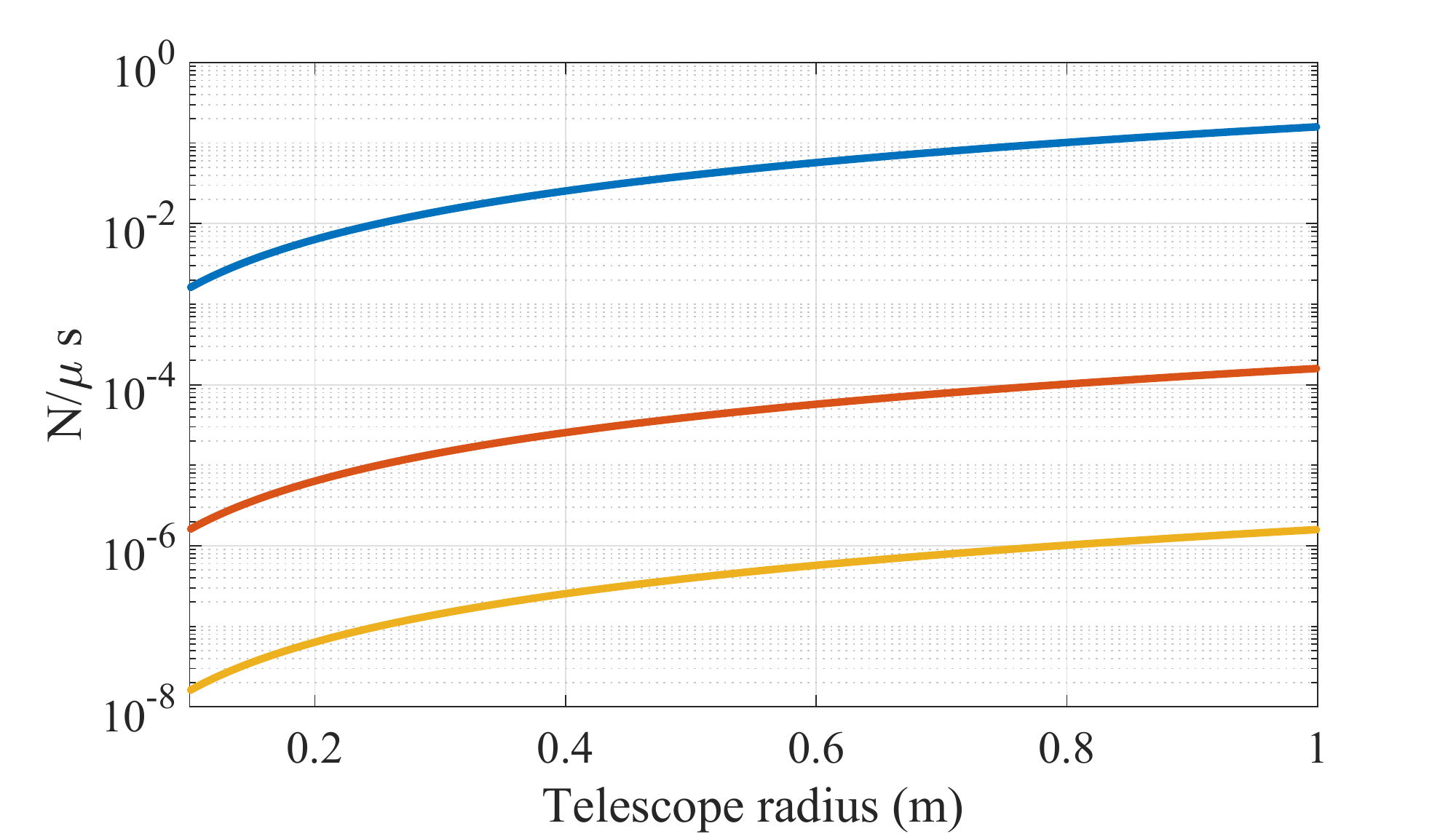}
    \caption{\label{fig:backgroundlight} Atmospheric background photons per $\mu s$ detection window. blue: $k=10^{-2}$ (clear daytime), red: $k=10^{-5}$ (full Moon night) and orange: $k=10^{-7}$ (moonless night).}
\end{figure}

The background photon rate can be calculated by following the approach in Ref.~\cite{Er_long2005}
to be
\begin{equation}
    R_{\mathrm{b}}= (\mathrm{k} \times H_{\mathrm{b}} \times \Omega_{\mathrm{fov}} \times A_{\mathrm{rec}} \times B_{\mathrm{filter}})/(h \times \nu),
\end{equation}
where $H_{\mathrm{b}}$ is the brightness of the sky (150~$\mathrm{W}\mathrm{m}^{-2}\mathrm{Sr}^{-1} \mathrm{\mu m}^{-1}$); $\Omega_{\mathrm{fov}}=3.14 \times 10^{-10}~\mathrm{Sr}$ is the field of view of the receiver telescope, which corresponds to an opening half-angle of $10\mathrm{\mu rad}$; $A_{\mathrm{rec}} = \pi R^2_{\mathrm{receiver}}$ is the area of the receiver telescope; $B_{\mathrm{filter}} = 0.02~\mathrm{nm}$ is the bandwidth of the spectral filter ($\sim$10~GHz) and finally $h$ is the Planck constant and $\nu = c/\lambda$ is the light frequency (with wavelength $\lambda = 780~\mathrm{nm}$). $k$ is a factor to account for different weather conditions where $k=10^{-2}$ corresponds to clear daytime and $k=10^{-5}$ and $k=10^{-7}$ correspond to clear night skies with full moon and no moon, respectively. We should note that in the case of satellite QKD, the relative motion between the satellite and the ground station creates a Doppler shift in the detected signal in the order of around $\sim10$~GHz which can be compensated with a tunable Fabry-Perot filter. In Fig. \ref{fig:backgroundlight} we plot the received noise photons per $1~\mu$s detection window as a function of telescope aperture for different atmospheric conditions. We see that operating during a moonless night with a large aperture telescope is comparable to the dark count rate of the detectors.
For the results in the main text we take $k=10^{-7}$, but some additional results for $k=10^{-5}$ can be found in Appendix \ref{sec:bright_night} where the influence of the background light has a more significant.

\subsection{Memory errors}

Each qubit that is held in memory is affected by time-dependent noise. We model this noise as a dephasing noise channel given by
\begin{equation}
    \mathcal{E}^{(i)}_z(t) \rho = (1-\lambda(t)) \rho + \lambda(t) Z^{(i)} \rho Z^{(i)},
\end{equation}
where
\begin{equation}
\lambda(t) = \frac{1 - e^{-t/T_\mathrm{dp}}}{2}
\end{equation}
with a memory specific dephasing time $T_\mathrm{dp}$.

\subsection{A note on quantum memories}
In this work, we focus on estimating the possible performance based on \textit{plausible} experimental parameters; however, we do not model a particular experimental setup down to its individual components. Nevertheless, our protocols require some advanced capabilities of the components involved - especially the quantum memories.

We assume that the memory modes of the multi-mode memories are individually addressable, so that it is possible, for example, to discard stored qubits where the other half of the pair has been lost in transit. Such spatial multi-mode operation with individual addressability was demonstrated with great success in \cite{Pu2017, Li2021}.

Furthermore, we assume that we can reliably tell when a qubit has been successfully loaded into memory via a quantum non-demolition (QND) measurement, which may be possible by using the same physical system as the quantum memories.
For example, in Ref. ~\cite{Sinclair2016}, QND detection of individual qubits was proposed and demonstrated using a quantum memory system based on rare earth materials.
Ref.~\cite{Rispe2011}, on the other hand, proposed QND detection of single photons at rubidium wavelength using an Rb-Bose-Einstein condensate, which is itself a promising quantum storage platform.
Single Rb atoms trapped in optical cavities have been used to demonstrate QND detection with efficiencies up to $74\%$ \cite{Reiserer2013}. In addition to atom-based QND schemes, it has been shown that purely photonic approaches can demonstrate heralded amplification of photonic qubits with overall efficiencies of about $30\%$ \cite{Bruno2014}.

\subsection{Possible extensions}
We do not model the following sources of imperfections, but they could be a starting point to expand the error model to include more details.
\begin{itemize}
    \item The initially generated entangled states may be imperfect.
    \item Bell measurements for entanglement swapping may also introduce additional errors.
    \item Experimentally other detectors may also be needed for certain processes, e.g., the heralding process when loading a photon into memory. These would also be subject to dark counts.
    \item For the most extreme distances, the satellite-satellite connections dip back into the atmosphere, which would likely limit the maximum achievable distance even before the horizon is reached.
\end{itemize}

\section{\label{sec:protocols}Protocols}
In the following we will give a detailed description of the protocols we have used.

\subsection{Scenario 1}

While the basic idea of the protocol is very similar to the one-satellite case and putting the sources on the satellites primarily serves to extend the range, there are some subtle timing differences that need to be taken into account.

The source satellites do not know whether the entangled pairs they send out have arrived successfully. Therefore, in this case it makes sense for the source to continuously send out pairs without waiting for confirmation from the other stations. Hence, the maximum rate of entangled pair generation $f_\mathrm{clock} = 20\text{ MHz}$ becomes an important parameter. We also use this clock rate for comparison with a setup without quantum memory. The clock rate is mainly limited by the memory bandwidth (see \cite{Gundogan2021} for an overview of suitable systems for quantum memories in space).

Here, we assume that the source continuously emits entangled pairs in time steps of $T_p = 1 / f_\mathrm{clock}$ and the memory has a mechanism to simply not accept any further qubits if a qubit is already in memory.

For the following derivation, consider a quantum memory with only one memory mode. The time for a pair to successfully arrive in memory is given by $T_p \times k(p_\mathrm{mem})$, where $k$ is sampled from a geometric distribution with probability $p_\mathrm{mem}$, which is the combined probability that the qubit arrives at $S_C$ and is successfully loaded into memory. After a qubit is loaded into memory, the memory must wait for $t_\mathrm{mem}$ for a message from the ground station as to whether the other qubit of the pair has arrived there. We have
\begin{equation}
 t_\mathrm{mem} = \frac{d_{S_AA} - d_{S_AS_C} + d_{AS_C}}{c},
\end{equation}
where $d_{IJ}$ denotes the distance between stations $I$ and $J$ and $c$ is the speed of light.

Of course, there is no guarantee that the other half of the pair will arrive successfully, so this must be repeated $j(p_A)$ times, where $j$ is sampled from a geometric distribution with probability $p_A$, which is the combined probability that a qubit will arrive at the ground station and be detected. Thus, the total time it takes to successfully establish a pair is given by
\begin{equation}
    j t_\mathrm{mem} + \sum_{i=1}^{j} k_i T_p,
\end{equation}
where each of the $k_i$ is an independent sample from the geometric distribution $k(p_\mathrm{mem})$.

For a quantum memory with $n$ modes, we assume that each time window is associated with a particular mode, so the net effect is that $n$ copies of the process described above run in parallel at a rate of $f_\mathrm{clock} / n$. Performance could likely be improved if the quantum memory is capable of dynamically selecting one of the empty modes each time a photon arrives.

\begin{figure}
    \centering
    \includegraphics[width=\linewidth]{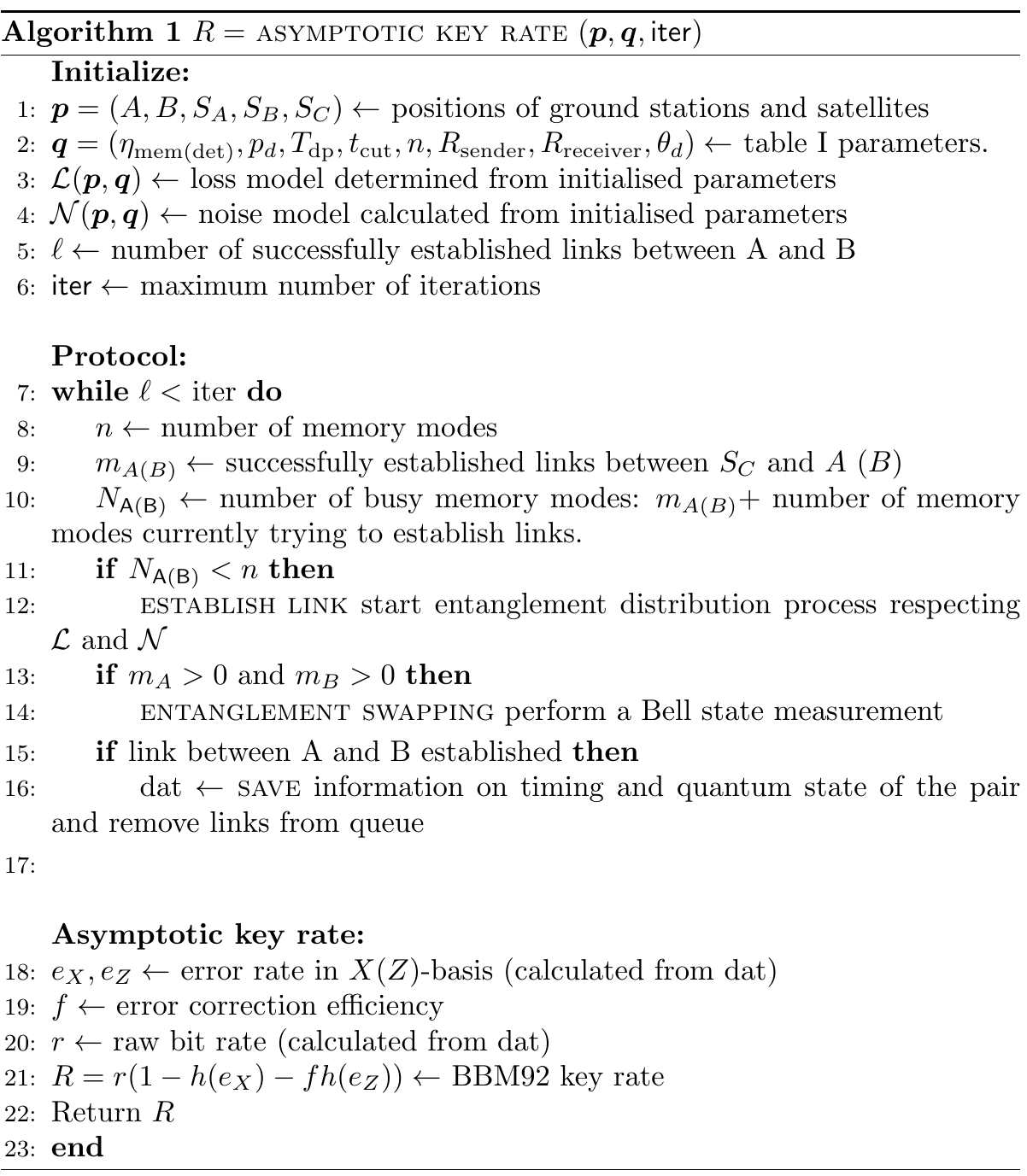}
    \caption{Pseudocode for the key rate for two ground stations $A$ and $B$ with three satellites (Scenario 1). We denote the positions of the ground stations and the satellites via the tuple $\bm{p}$, all parameters in table 1 (main body of paper) via the tuple $\bm{q}$, and $\mathsf{iter}$ as the number of iterations. The code then determines a loss and noise model from $\{\bm{p}, \bm{q}, \mathsf{iter}\}$ and returns the achievable rate.}
    \label{fig:pseudocode}
\end{figure}%

In Fig.~\ref{fig:pseudocode} the simulation procedure to obtain the key rate for Scenario 1 is illustrated. This requires initialisation of all system parameters listed in Table 1 in the main body of the paper. The loss and noise model is determined from the initialised parameters in accordance to section~\ref{sec:error_models}. The pseudocode then executes the protocol by establishing links and performing entanglement swapping between available links. It accounts for the subtle timing difference between busy memory modes and those trying to establish a link. All information on timing and the final quantum state of successfully linked memory modes are saved. This information is then used to determine the asymptotic key rate using the BBM92 protocol. For increasing number of satellites, the protocol can be extended by introducing a similar logic of establishing repeater links for each entangled pair source and a procedure for entanglement swapping at the satellites with quantum memories.

\subsection{Scenario 2}

The satellites with the capability to generate entangled states, $S_A$ and $S_B$, try to establish pairs with both their associated ground station and the central satellite.
For now, let us again consider quantum memories with just one memory mode. The emissive quantum memory generates an entangled memory-photon pair. The photon is then sent through a channel toward its destination. The qubit at the satellite must remain in memory until a message can be received from the ground station as to whether the transmitted qubit arrived successfully.

This means for establishing a link with a ground station, it takes $t^A_\mathrm{trial} = 2 d_{S_AA} / c$ for one trial (and equivalent for $B, S_B$). This also means that the qubit at $S_A$ will have been stored in memory (and therefore been dephased) for $t^{A}_{\mathrm{trial}}$ even if the trial has been successful. Similarly, when establishing a link between $S_A$ and $S_C$ one trial will take $t^{A^\prime}_\mathrm{trial} = 2 d_{S_A S_C} / c$. For a successfully established pair the qubit at $S_A$ will have dephased for $t^{A^\prime}_\mathrm{trial}$ while the qubit at $S_C$ will have dephased for $t^{A^\prime}_\mathrm{trial} / 2$.

Similar to Scenario 1 we can calculate the number of trials necessary to establish a link $k$ by drawing randomly from a geometric distribution $k(p)$, where $p$ is the probability that a pair will be established in one trial (i.e. the combined probability that the qubit arrrives and both qubits are loaded into memory successfully). Therefore, the total time it takes until the next pair is established is given by
$k t_\mathrm{trial}$
with a $p$ and $t_\mathrm{trial}$ specific to the considered link.

In general we consider multi-mode memories and assume that the above process can be done in parallel and independently for each mode. In practice, it is likely that a spatially separate channel will not be available for each memory mode, but instead they would need to use specific time slots of a shared channel. However, if the number of distinguishable time slots for the channel (which we directly link to $f_\mathrm{clock}$ for the purposes of the simulation) is much larger than the number of memory modes, the effects of sharing a channel are negligible. This is in contrast to setups with no quantum memory being utilized at the source satellites (like Scenario 1), where the capacity of the channel is indeed a relevant limitation.

When two qubits corresponding to successful links on each side are in memory, the satellite performs a Bell state measurement and communicates the result to the neighboring stations. If there are multiple eligible qubits available on one side, the qubit that has been in memory the longest is used. While it certainly could be beneficial to use pairs that have not been dephased as much, this choice avoids satellites having to exchange additional classical messages about their strategy.
More advanced schemes would need to coordinate which qubits to use in the Bell measurement for entanglement swapping. This provides an additional challenge in formulating a protocol, since e.g.~two satellites might need to make a decision simultaneously. This could either lead to cases where they make incompatible decisions and therefore achieve a lower overall efficiency or one would need to introduce some kind of time slots in which a specific satellite is allowed to do operations to make sure all relevant information is available, which would lead to qubits that stay in memory for longer. Therefore, for this work we limited ourselves to a simple protocol without the need for additional communication.

\subsection{One satellite}
For comparison with our three satellites protocol we also investigated setups using only one satellite, while keeping the noise models and parameters the same.
For a single satellite with quantum memories we use a protocol that works exactly the same as Scenario 2 described above, with the exception of there being only two links instead of four: One link between the satellite and $A$ and one link between the satellite and $B$. This is very similar to the downlink protocol in Ref.~\cite{Gundogan2021}. However, here we use a slightly different strategy that allows for simultaneous trials in both directions and utilizing cutoff times, whereas Ref.~\cite{Gundogan2021} uses the protocol from Ref.~\cite{Trenyi2020}.

\begin{figure*}
    \centering
    \includegraphics[width=0.9\linewidth]{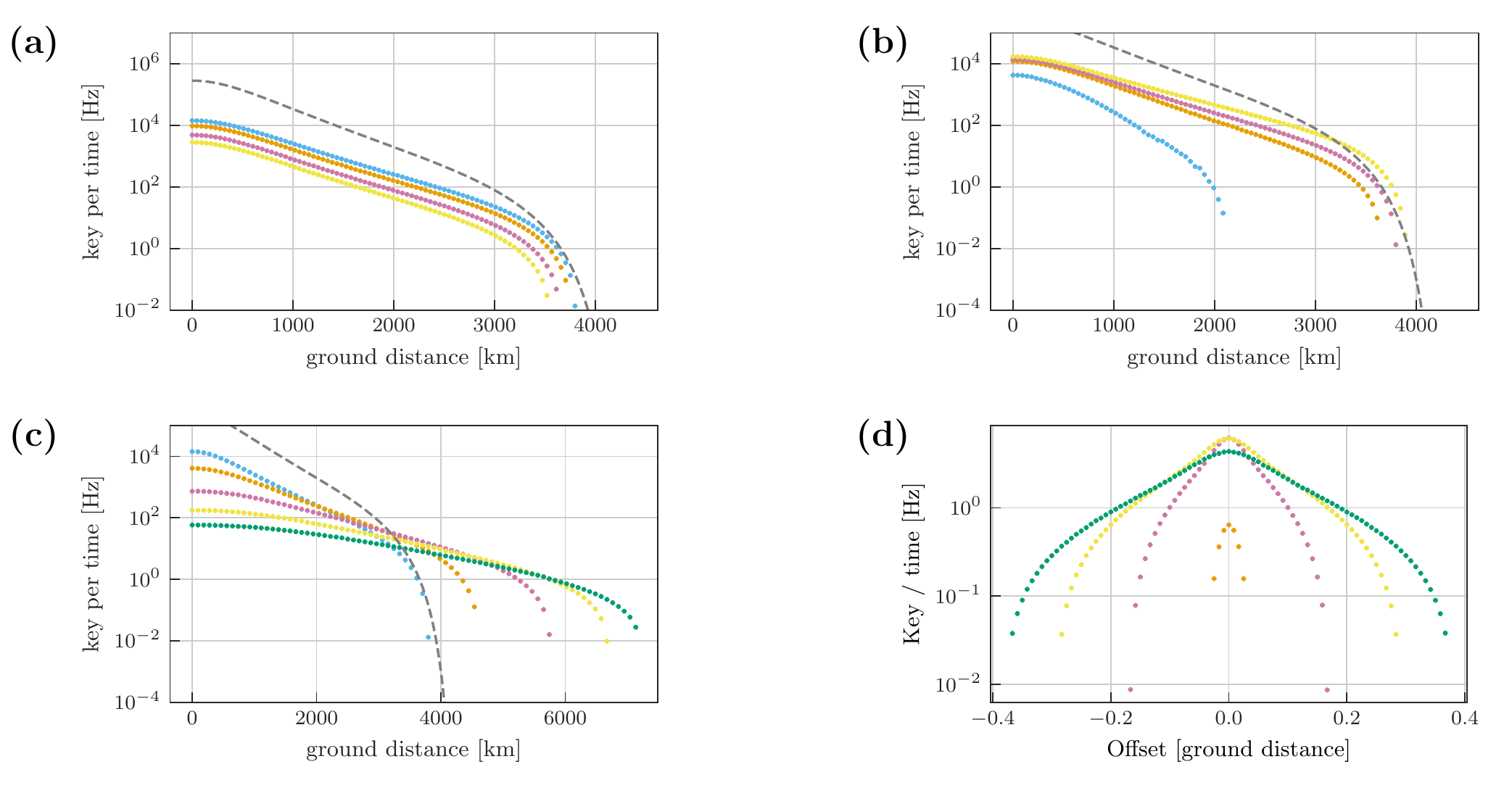}
    \caption{Achievable key rates for one satellite with quantum memories. The dashed line indicates an idealized one satellite repeater protocol without memories. (a) Different divergence angles $\theta_d=$ $3\mathrm{\mu rad}$ (blue), $4\mathrm{\mu rad}$ (orange), $6\mathrm{\mu rad}$ (pink), and $8\mathrm{\mu rad}$ (yellow). (b) Varying memory qualities $T_\mathrm{dp}=$ $10 \mathrm{ms}$ (blue), $50 \mathrm{ms}$ (orange), $100 \mathrm{ms}$ (pink), $1000 \mathrm{ms}$ (yellow). (c/d) Varying orbital heights $h=$ $400\mathrm{km}$ (blue), $600\mathrm{km}$ (orange), $1000\mathrm{km}$ (pink), $1500\mathrm{km}$ (yellow), $2000\mathrm{km}$ (green).}
    \label{fig:one_satellite}
\end{figure*}

Additional results from this one satellite scenario are depicted in Fig.~\ref{fig:one_satellite}. The key result from this scenario is that while increasing the orbital height does indeed help to reach longer distances, for our parameter set the key rates go down drastically due to the additional diffraction loss and provide significantly lower rates than the best satellite configurations in Scenario 1.

It should be noted that an uplink protocol with entangled pair sources located at the ground stations instead of in spaces would also be a possible choice of protocol. Repeater protocols with one intermediate station that has the end stations send pairs to the central station can be very efficient, see, 
e.g., Ref.~\cite{QLinkXwhitepaper} for a non-satellite example. However, for satellites uplink protocols have the additional complication significant additional divergence as the laser beam travels through the atmosphere first. Despite this limitation the possibility of the ground stations actively sending entangled photons should be kept in mind as an option for future investigations.

As a last setup, we also consider a single satellite without memory that simply generates pairs and immediately sends one half of each pair to $A$ and the other half to $B$. For a quick estimate how this would perform, we take into account all sources of loss as with he full model above, but omit all the effects that would lead to loss in fidelity, e.g., dark counts and background light. So the performance of this baseline, to which we compare our other scenarios, is actually slightly overestimated.

\section{\label{sec:estimate} Estimate of dynamic key rates from static results}
As mentioned in the main text we let our simulation run until a large number of bits have been obtained and use this sample to calculate the key rates. The output of the simulation is essentially a list of times the successful trials happened and the associated quantum state. We then obtain the sample mean of $e_X$ and $e_Z$ (error rate for measurements in $X$ and $Z$ direction, respectively)  from all the density matrices, as well as the raw rate $r$, which is simply the number of distributed bits divided by the total time it took to distribute them.

Now consider multiple runs of the simulation for satellite positions along the path of the satellite, e.g., Fig.~\ref{fig:satellite_path}b. For each run we perform the above calculations and denote them as $e_{X(Z)}(t)$ and $r(t)$. We choose $t=0$ to be the point in the orbit where the satellites are positioned exactly between the two ground stations.

The number of raw bits obtained from one pass of the satellits is given by
\begin{equation}
    \bar{R} = \int_{-\tau}^{\tau} r(t) \mathrm{d}t ,
\end{equation}
and the average error rates of the bit strings arriving at the ground station can be calculated to
\begin{equation}
    \bar{e}_{X(Z)} = \frac{\int_{-\tau}^{\tau} r(t) e_{X(Z)}(t) \mathrm{d}t}{\bar{R}}.
\end{equation}
$\tau$ is a chosen position in the orbit beyond which the bits arriving will not be used. This is important because while there may still be signals arriving from the satellites, at some point the fidelity of the distributed pair of qubits may become too low to be useful. We choose $\tau$ such that the obtainable key rate is optimized. Since we only have a limited number of data points available, we perform the integration numerically by the trapezoidal rule.

The orbital period of a satellite at orbital height $h$ is given by
\begin{equation}
    T(h) = 2 \pi \sqrt{\frac{(R_E + h)^3}{M G}}
\end{equation}
with the average earth radius $R_E = 6371 \ \mathrm{km}$, the earth mass $M = 5.972 \times 10^{24} \ \mathrm{kg}$ and the gravitational constant $G = 6.67408 \times 10^{-11} \ \mathrm{m}^3 \mathrm{kg}^{-1} \mathrm{s}^{-2}$.

With that we can calculate the effective key rate (i.e., also taking into account the time spent waiting for the satellites being reachable again)
to be
\begin{equation}
    \frac{R}{T(h)} \left(1 - h(\bar{e}_X) - f h(\bar(e)_Z) \right).
\end{equation}
This process assumes that all the raw bits obtained over one or multiple satellite passes are pooled together and the additional information that some bits are more likely to contain errors than others is not used.
5
The results for some of the satellite configurations have been presented in Table \ref{tab:satellite_path_results}. In Table \ref{tab:satellite_path_full}, the effective key rates for all the scenarios we simulated are listed.

\begin{table*}[]
    \centering
    \begin{tabular}{c|c|c|c|c}
        Configuration & raw bits per pass & raw bits per second & key bits per pass & key bits per second \\
        \hline
         Scenario 1, $S_A @ 0\%$ & $1.1 \times 10^6$ & $2.0 \times 10^2$ & $5.3 \times 10^5$ & $9.6 \times 10^1$ \\
         Scenario 1, $S_A @ 10\%$ & $3.9 \times 10^5$ & $7.1 \times 10^1$ & $6.1 \times 10^4$ & $1.1 \times 10^1$ \\
         Scenario 1, $S_A @ 20\%$ & $9.2 \times 10^4$ & $1.7 \times 10^1$ & $7.9 \times 10^3$ & $1.4 \times 10^0$ \\
         Scenario 1, $S_A @ {-10}\%$ & $2.8 \times 10^5$ & $5.1 \times 10^1$ & $3.5 \times 10^4$ & $6.3 \times 10^0$ \\
         Scenario 2, $S_A @ 0\%$ & $1.1 \times 10^5$ & $2.0 \times 10^1$ & $1.4 \times 10^3$ & $2.5 \times 10^{-1}$ \\
         Scenario 2, $S_A @ 10\%$ & $1.4 \times 10^5$ & $2.6 \times 10^1$ & $2.3 \times 10^3$ & $4.1 \times 10^{-1}$ \\
         Scenario 2, $S_A @ 20\%$ & $1.1 \times 10^5$ & $2.0 \times 10^1$ & $2.3 \times 10^3$ & $3.0 \times 10^{-1}$ \\
         Scenario 2, $S_A @ {-10}\%$ & $5.0 \times 10^4$ & $9.1 \times 10^0$ & $4.1 \times 10^2$ & $7.3 \times 10^{-2}$ \\
         1 satellite with memory, $h=600\mathrm{km}$ & $2.8 \times 10^2$ & $4.9 \times 10^{-2}$ & $2.9 \times 10^1$ & $5.1 \times 10^{-3}$ \\
         1 satellite with memory, $h=1000\mathrm{km}$ & $6.8 \times 10^3$ & $1.1 \times 10^0$ & $9.3 \times 10^2$ & $1.5 \times 10^{-1}$ \\
         1 satellite with memory, $h=1500\mathrm{km}$ & $1.3 \times 10^4$ & $1.9 \times 10^0$ & $1.7 \times 10^3$ & $2.4 \times 10^{-1}$ \\
         1 satellite with memory, $h=2000\mathrm{km}$ & $1.6 \times 10^4$ & $2.1 \times 10^0$ & $1.8 \times 10^3$ & $2.3 \times 10^{-1}$ \\
         1 satellite without memory, $h=600\mathrm{km}$ & & & $7.3 \times 10^1$ & $1.3 \times 10^{-2}$ \\
         1 satellite without memory, $h=1000\mathrm{km}$ & & & $7.7 \times 10^3$ & $1.2 \times 10^0$ \\
         1 satellite without memory, $h=1500\mathrm{km}$ & & & $2.2 \times 10^4$ & $3.2 \times 10^0$ \\
         1 satellite without memory, $h=2000\mathrm{km}$ & & & $2.7 \times 10^4$ & $3.5 \times 10^0$
    \end{tabular}
    \caption{Estimates for effective rates for three or one satellite orbiting earth.}
    \label{tab:satellite_path_full}
\end{table*}

\section{Additional results for full moon}
\label{sec:bright_night}
In Section \ref{sec:background_light} we discuss our model for background light leading to additional erroneous detector clicks.
For the base case of relative brightness $k=10^{-7}$ this has an effect similar to the dark counts $p_d$. However, let us consider the case of $k=10^{-5}$ which would be appropriate for a clear night with a full moon.

\begin{figure*}
    \centering
    \includegraphics[width=0.9\linewidth]{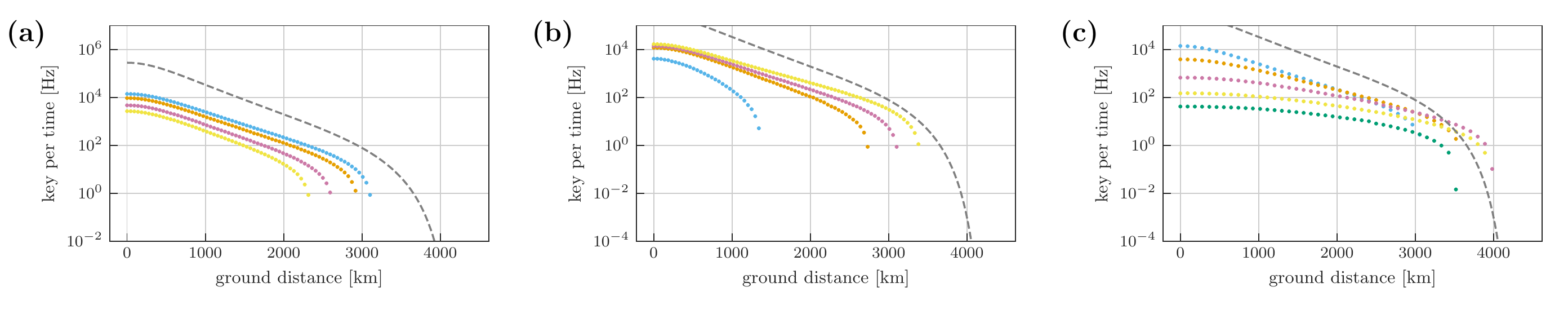}
    \caption{Achievable key rates for one satellite with quantum memories with significant background light (relative brightness $k=10^{-5}$). The dashed line indicates an idealized one satellite repeater protocol without memories. (a) Different divergence angles $\theta_d=$ $3\mathrm{\mu rad}$ (blue) $4\mathrm{\mu rad}$ (orange) $6\mathrm{\mu rad}$ (pink) and $8\mathrm{\mu rad}$ (yellow). (b) Varying memory qualities $T_\mathrm{dp}=$ $10 \mathrm{ms}$ (blue), $50 \mathrm{ms}$ (orange), $100 \mathrm{ms}$ (pink), $1000 \mathrm{ms}$ (yellow). (c/d) Varying orbital heights $h=$ $400\mathrm{km}$ (blue), $600\mathrm{km}$ (orange), $1000\mathrm{km}$ (pink), $1500\mathrm{km}$ (yellow), $2000\mathrm{km}$ (green).}
    \label{fig:bright_night_one_satellite}
\end{figure*}

\begin{figure*}
    \centering
    \includegraphics[width=0.9\linewidth]{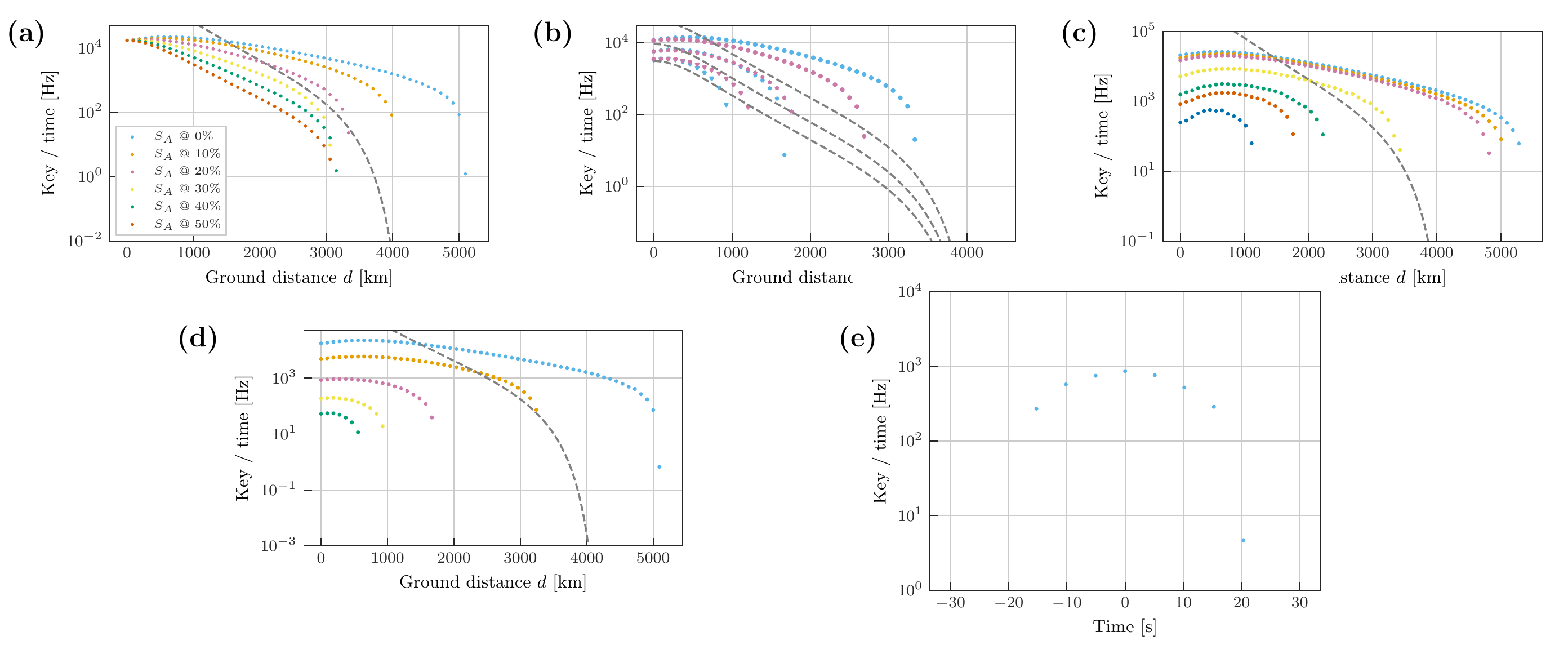}
    \caption{Scenario 1 with significant background light (relative brightness $k=10^{-5}$). In each subfigure one parameter is varied, while the others are kept at their base value. (a) Satellite $S_A$ is positioned vertically above varying percentages of the total ground distance. $S_A @$ $0\%$ (blue), $10\%$ (pink), $20\%$ (orange), $30\%$ (yellow), $40\%$ (green), $50\%$ (red) (b) Higher divergence angle $\theta_d = 4\mu\mathrm{rad}, 6\mu\mathrm{rad}, 8\mu\mathrm{rad}$ and therefore higher loss for $S_A$ satellite positions $0$ (blue) and $0.2$ (purple). (c) Various memory qualities with dephasing times $T_\mathrm{dp}$ of  $1\mathrm{s}$ (blue), $100\mathrm{ms}$ (orange), $50\mathrm{ms}$ (pink), $10\mathrm{ms}$ (yellow), $5\mathrm{ms}$ (green), $4\mathrm{ms}$ (red), $3\mathrm{ms}$ (dark blue), $2\mathrm{ms}$ (black). (d) Differing orbital heights for all three satellites $400\mathrm{km}$ (blue), $600\mathrm{km}$ (orange), $1000\mathrm{km}$ (pink), $1500\mathrm{km}$ (yellow), $2000\mathrm{km}$ (green). (e) For ground distance $d=4400\mathrm{km}$ only the the case with satellites spaced optimally (compare (a) ) can reach, but even so a positive key rate can only be obtained for a very small time window.
    }
    \label{fig:bright_night_twolink}
\end{figure*}

For the one satellite case the results are depicted in Fig.~\ref{fig:bright_night_one_satellite}. We also performed the same analysis for Scenario 1 with results in Fig. \ref{fig:bright_night_twolink}. While there still exists a slight advantage for using three satellites, the background light is very significant and severely limits the reachable distances.

\section{Essential software}
The code for the simulation is written in Python 3 \cite{python3_manual}, with the python packages NumPy \cite{numpy}, pandas \cite{pandas} and matplotlib \cite{matplotlib} being the core of our program.

\end{document}